\documentclass[12pt,a4paper]{article}

\usepackage{geometry}
\usepackage{hyperref}
\usepackage{cite}
\usepackage{url}
\usepackage{amsmath}
\usepackage{amssymb}
\usepackage[dvips]{graphicx}
\usepackage{latexsym}
\usepackage{pstricks}
\usepackage{bm}
\usepackage{pbox}
\usepackage{placeins}
\usepackage{graphicx}% Include figure files
\usepackage{caption}
\usepackage{subcaption}
\usepackage[T1]{fontenc}
\usepackage{footnote}
\usepackage{pdfpages}
\usepackage{hhline}
\usepackage{multirow}
\usepackage{enumitem}
\usepackage[toc,page]{appendix}
\newcommand{\be}{\begin{equation}}
\newcommand{\ee}{\end{equation}}

\newcommand{\bea}[1]{\begin{eqnarray}\label{#1} }
\newcommand{\eea}{\end{eqnarray}}

\newcommand\scalemath[2]{\scalebox{#1}{\mbox{\ensuremath{\displaystyle #2}}}}

\def\beqn{\begin{eqnarray}}
\def\eeqn{\end{eqnarray}}

\def\beq{\begin{equation}}
\def\eeq{\end{equation}}
\def\bea{\begin{eqnarray}}
\def\eea{\end{eqnarray}}

\def\vs{\vspace}

\newcommand{\bem}{\begin{pmatrix}}
\newcommand{\eem}{\end{pmatrix}}

\begin{document}
\vspace*{-0.2in}
\begin{flushright}
OSU-HEP-16-08\\
%hep-ph/
\end{flushright}

\vs{0.5cm}

\renewcommand{\thefootnote}{\fnsymbol{footnote}}

\begin{center}
{\Large\bf Yukawa Sector of Minimal \boldmath{$SO(10)$} Unification }\\
\end{center}

\vspace{0.5cm}
\begin{center}
{\large
{}~K.S. Babu$^{a,}$\footnote{E-mail: babu@okstate.edu},{}~
Borut Bajc$^{b,}$\footnote{E-mail: borut.bajc@ijs.si} and
{}~Shaikh Saad$^{a,}$\footnote{E-mail: shaikh.saad@okstate.edu}
}
\vspace{0.5cm}

\centerline{$^{a}${\it\small Department of Physics, Oklahoma State University, Stillwater, OK, 74078, USA }}
\centerline{$^{b}${\it\small Jo\v{z}ef Stefan Institute, 1000 Ljubljana, Slovenia}}
\end{center}
%\vspace{0.6cm}

%%%%%%%%%%%%%%%%%%%%%%%%%%%%%%%%%%%%%%%%%%%%%
%%%%%%%%%%%%%%%%%%%%%%%%%%%%%%%%%%%%%%%%%%%%%
\begin{abstract}
We show that in $SO(10)$ models, a Yukawa sector consisting of a real $10_H$, a real $120_H$ and a complex $126_H$
of Higgs fields can provide a realistic fit to all fermion masses and mixings, including the neutrino sector.  Although the
group theory of $SO(10)$ demands that the $10_H$ and $120_H$ be real, most constructions complexify these fields
and impose symmetries exterior to  $SO(10)$ to achieve predictivity.  The proposed new framework with {\it real} $10_H$
and {\it real} $120_H$ relies only on $SO(10)$ gauge symmetry, and yet has a limited number of Yukawa parameters.
Our analysis shows that while there are restrictions on the observables, a good fit to the entire fermion spectrum can
be realized. Unification of gauge couplings is achieved with an intermediate scale Pati-Salam gauge symmetry.
Proton decay branching ratios are calculable, with the leading decay modes being
$p \rightarrow \overline{\nu} \pi^+$ and $p \rightarrow e^+ \pi^0$.
\end{abstract}

\renewcommand{\thefootnote}{\arabic{footnote}}
\setcounter{footnote}{0}
\thispagestyle{empty}

\newpage
%%%%%%%%%%%%%%%%%%%%%%%%%%%%%%%%%%%%%%%%%%%%%
%%%%%%%%%%%%%%%%%%%%%%%%%%%%%%%%%%%%%%%%%%%%%
\section{Introduction}

Grand unified theories (GUTs) \cite{Pati:1974yy,Georgi:1974sy,Georgi:1974yf} based on the  gauge group
$SO(10)$ \cite{so10} are very attractive candidates to unify the strong, weak and electromagnetic forces into a single force, as well as to
shed light on some of the open questions of  the Standard Model (SM).  Quarks and leptons of each family
are unified into a single irreducible representations of $SO(10)$ group, the 16-dimensional spinor, which also contains the right-handed
neutrino.  The presence of the right-handed neutrino makes the seesaw mechanism \cite{Minkowski:1977sc,Schechter:1980gr} for generating small neutrino masses very compelling in these theories.  Since $SO(10)$ gauge symmetry is automatically anomaly-free \cite{so10}, it provides a nice explanation for the miraculous cancelation of anomalies that occurs within each fermion family.  The observed quantization of electric charges is also understood in these theories owing to their non-Abelian nature.  Unifying all fermions into a single multiplet gives us the hope of understanding some aspects of the flavor puzzle in these theories.  Unification of gauge couplings occurs naturally
at an energy scale of $\sim 10^{15-16}$ GeV \cite{GoranIntSc:1981,Caswell:1982fx,Chang:1983fu,Gipson:1984aj,Chang:1984qr,Deshpande:1992au,Deshpande:1992em,Bertolini:2009qj,Bertolini:2009es,
Bertolini:2010ng,khan,Graf:2016znk}, as $SO(10)$ admits an intermediate symmetry group -- unlike theories based on $SU(5)$ which must break directly to the SM.  It is of course well known that if supersymmetry is assumed to be present in its minimal version at the TeV scale, one-step breaking of $SO(10)$ directly down to the SM can be realized at an energy scale of $2\times 10^{16}$ GeV \cite{susyunification}. The focus of this paper is, however, $SO(10)$ theories without the assumption of supersymmetry.

We wish to inquire what an economic Yukawa sector would look like in renormalizable $SO(10)$ theories.
This may appear to be a well understood issue, but as we suggest here, this question has not been properly resolved.
Economy may be viewed as having the least number of Higgs fields as well as Yukawa parameters while being realistic.  Assuming that there are no new fermions beyond the three families of chiral 16s\footnote{If vector-like fermions belonging to $16 + \overline{16}$ (or other real representations) with GUT scale masses exist and  mix with the chiral 16s, new possibilities are available, see for e.g., Ref. \cite{bb,bbs}.} the answer to this question may be found in the group theory of fermion bilinears:
%\vspace*{-5pt}
%\begin{align}
\begin{equation}
16\times 16 = 10_s+ 120_a + 126_s.
\end{equation}
%\end{align}
%\vspace*{-5pt}
Here the subscripts $s$ and $a$ stand for symmetric and antisymmetric components (in family space).  The 10 and the 120 are real representations in $SO(10)$, while the 126 is complex.  The most general renormalizable Yukawa couplings in $SO(10)$ theories then would take the form
\begin{equation}
\label{yukawa0}
\mathcal{L}_{yuk}= 16_F (Y_{10}^i10_H^i+Y_{120}^j 120_H^j+Y_{126}^k\overline{126}_H^k) 16_F.
\end{equation}
Here the index $i$ takes values $i= 1, 2,..n_{10}$ where $n_{10}$ is the number of $10_H$ fields employed, and similarly the index $j = 1, 2, ..n_{120}$ and $k = 1, 2, ..n_{126}$ with $n_{120}$ and $n_{126}$ being the number of $120_H$ and $\overline{126}_H$ present in the theory. The Yukawa coupling matrices $Y_{10}^i$ and $Y_{126}^k$  are $3 \times 3$ complex  symmetric matrices in family space, while $Y_{120}^j$
are complex antisymmetric matrices.  We wish to identify the smallest possible set of \{$n_{10}, \, n_{120},\, n_{126}$\} that would lead to a realistic spectrum of quark and lepton masses as well as mixing angles. This set will turn out to be the choice $n_{10} = n_{120} = n_{126} = 1$, as we shall see. This result is satisfying, as it suggests that nature has utilized each possible Higgs field for fermion mass generation exactly once, without any replication.

Before establishing this assertion, which will be done in the next section, let us note that {\it  a complex} $10$ can be constructed from {\it two real} 10s in $SO(10$):  $10_c = (10_1 + i 10_2)/\sqrt{2}$. Similarly, {\it a complex} $120_c$ may be constructed from {\it two real} 120s.  In these cases, the Yukawa couplings will involve terms of the type $16_F 10_c 16_F$ as well as $16_F 10_c^* 16_F$ with completely independent Yukawa coupling matrices, and similarly for the $120_c$ field.  It is possible to assign a charge exterior to $SO(10)$ to these fields -- such as the Peccei--Quinn $U(1)$ motivated on other grounds -- so that the Yukawa couplings contain only the $16_F 10_c 16_F$ term, and not the $16_F 10_c^* 16_F$ term.  These restricted class of Yukawa couplings in $SO(10)$ have been studied extensively \cite{Davidson:1983fe,Lazarides:1980nt,Lazarides:1990ni,Babu:1992ia,Lee:1994qx,Bajc:2005zf,Joshipura:2011nn,Altarelli:2013aqa,Dueck:2013gca}.  While interesting, the predictions of such models are those of $SO(10) \times G$ where $G$ is a symmetry exterior to $SO(10)$, and not of the true grand unified symmetry $SO(10)$ itself. Our inquiry relates to the minimal Yukawa sector in theories where only the $SO(10)$ gauge symmetry plays a role.

It should be noted that in theories which assume supersymmetry (SUSY), which is not the focus of the present work, chiral superfields are necessarily complex, thus requiring the complexification of 10 and 120 Higgs fields.  Holomorphy of the superpotential would imply that the coupling $16_F 10_c^* 16_F$ is not present simultaneously with the superpotential term $16_F 10_c16_F$.  These models share some of the features of non-SUSY models based on $SO(10) \times U(1)_{PQ}$, although the renormalization group evolution of the fermion mass parameters between the weak scale and the GUT scale would be different in the two classes of theories.  Supersymmetric $SO(10)$ models have been studied extensively, and it has been shown that economic models where only a (complex) $10_H$ and a $\overline{126}_H$ couple to fermions can be predictive and consistent with all fermion masses and mixings \cite{Babu:1992ia,Bajc:2001fe,Fukuyama:2002ch,Bajc:2002iw,Goh:2003sy,Goh:2003hf,Babu:2005ia,Bertolini:2006pe,Bajc:2008dc,Fukuyama:2015kra} \cite{Joshipura:2011nn,Dueck:2013gca}.  If the additional Higgs fields needed for symmetry breaking are restricted to a $126_H$ and a $210_H$, split supersymmetry may be required for consistency \cite{Bajc:2008dc}.  Alternatively, a (complex) $120_H$ may be introduced to relax some of the restrictions imposed by the symmetry breaking sector \cite{Bertolini:2004eq,Yang:2004xt,Dutta:2004zh,Aulakh:2008sn}.\footnote{Symmetries external to $SO(10)$ have also been applied in the context of renormalizable SUSY $SO(10)$ with some success in explaining the fermion spectrum.  See for example Ref. \cite{
Chen:2003zv,Grimus:2006bb,Cai:2006mf,Albaid:2011vr,Ferreira:2015jpa}.}
Our goal in this paper is to identify the analog of the minimal SUSY $SO(10)$ Yukawa sector, but for $SO(10)$ theories without supersymmetry.

The rest of the paper is organized as follows.  In Sec. 2 we present our proof that the economic Higgs sector will have $n_{10} = n_{120} = n_{126} = 1$.  In Sec. 3 we analyze the predictions of this model for quark and lepton masses and mixings.  Here we present our numerical study which shows full consistency with experimental data.  In Sec. 4 we present the constraints on these models from the unification of gauge couplings; in Sec. 5 we calculate the proton decay branching ratios. In Sec. 6 we conclude. Four Appendices contain a discussion on how the fine-tuning is achieved, technical details on the proof of economic Yukawa sector as well as the best fit parameters for the minimal model.

\section{Economic Yukawa Sector in \boldmath{$SO(10)$}}

In this section we establish the assertion that $n_{10} = n_{120} = n_{126} = 1$ is the economic choice of Yukawa sector in non-supersymmetric $SO(10)$ theories.  This corresponds to choosing one real $10_H$, one real $120_H$ and a complex $\overline{126}_H$ of Higgs bosons that have Yukawa couplings with the three chiral families of $16_F$. An additional Higgs filed belonging to $45_H$, $54_H$ or $210_H$ would be needed for completing the symmetry breaking.  These fields, however, do not have Yukawa couplings with the $16_F$, and the precise choice is not so important for now.  A proof of our assertion would require that the choice $n_{10} = n_{120} = n_{126} = 1$ leads to a realistic fermion spectrum, and no other simpler choice exists consistent with realism.  The former part of the proof is delegated to Sec. 3 where we perform a numerical analysis of this economic Yukawa sector; here we address the latter part.

If only one Higgs field among $10_H$, $120_H$ and $\overline{126}_H$ is present in a theory, there would be no flavor mixing -- as the Yukawa coupling matrix of this single Higgs field can be diagonalized using an $SO(10)$ rotation.  Thus at least two Higgs fields are needed for realistic fermion spectrum.  One of the fields used must be a $126_H$, since it gives large Majorana masses to the righ-handed neutrinos directly.  This field also plays a role in the symmetry breaking sector, as it breaks $SO(10)$ down to $SU(5)$, reducing the rank.  One could consider replacing the $126_H$ with a $16_H$ which can play a similar role in rank reduction.  In such a case the right-handed neutrino can acquire a large Majorana mass via the two-loop Witten diagram \cite{witten} involving gauge boson and scalar loops.  The induced Majorana mass can be estimated \cite{bs} to be of order
\begin{equation}
M_{\nu^c} \approx \left(\frac{\alpha_{10}}{4 \pi}\right)^2 Y_{10} \frac{v_R^2}{M_{\rm GUT}}
\label{two-loop}
\end{equation}
where $\alpha_{10}$ is the $SO(10)$ gauge coupling, $Y_{10}$ is the Yukawa coupling of $10_H$, and $v_R$ is the $B-L$ breaking vacuum expectation value (VEV) of the $16_H$.  In a nonsupersymmetric $SO(10)$ theory $v_R$ is well below the GUT scale for consistency with gauge coupling unification, with its range being $v_R \approx (10^{11} - 10^{14})$ GeV depending on the surviving intermediate symmetry.  $M_{\nu^c}$ is then of order $10^8$ GeV or less, which is too small to reproduce the correct order of magnitude for the light neutrino masses.\footnote{This issue with the Witten mechanism may be resolved in split supersymmetry, where $v_R = M_{\rm GUT}$ \cite{bs}. The SUSY particle masses should be of order the GUT scale to prevent additional suppression factor of $M_{\rm SUSY}/M_{\rm GUT}$ in Eq. (\ref{two-loop}), which may cause a problem with generating a Higgs boson mass of 125 GeV \cite{Giudice:2011cg}.}

Keeping one $126_H$ field in the theory, we seek if a realistic fermion spectrum can be generated with the addition of a second Higgs field.  This turns out to be not possible.  If the second Higgs field is a $126_H$, the mass relations $m_\tau = -3 m_b, m_\mu = -3 m_s$ and  $m_e = -3 m_d$ will result at the GUT scale, which are inconsistent with observations.  The ratio $m_\tau/m_b$ is found to be about 1.7 at the GUT scale (with small input errors) when the low energy mass parameters are evolved up to the GUT scale using SM renormalization group equations.  We found that this ratio is more realistically in the range $(1.4-1.7)$, when intermediate scale threshold effects arising from the right-handed neutrino sector and the gauge bosons of $SU(4)_c$ are included.  Each of the two threshold effects causes a decrease in the ratio $m_\tau/m_b$ at the GUT scale.  We conclude that the relation $m_\tau = 3 m_b$ is clearly excluded.  The relation $m_\mu = 3 m_s$ is not too far off (our RGE evolution shows the ratio $m_\mu/m_s$ to be about 4 at the GUT scale), while $m_e = 3 m_d$ is off by an order of magnitude. Thus a minimal Yukawa sector consisting of two copies of $\overline{126}_H$ is not realistic.

If the second Higgs field is a real $10_H$, two complex symmetric Yukawa matrices can be written down, one with the $10_H$, and one with the $\overline{126}_H$.  However, the Higgs doublet in the $10_H$ is self-conjugate, and is contained in the $(2,2,1)$ representation of the Pati-Salam subgroup $SU(2)_L \times SU(2)_R \times SU(4)_c$.  This field can  be written as
\begin{eqnarray}
\Phi^{\ast} = \tau_2 \Phi \tau_2 \Rightarrow \Phi= \begin{pmatrix}
\phi_0&\phi^+\\
-\phi^-&\phi^{\ast}_0
\end{pmatrix}.
\label{real}
\end{eqnarray}
In general, if the (1,1) element of $\Phi$ is independent from the (2,2) element, we can denote their respective vacuum expectation values to be $v_u$ and $v_d$ with $v_u$ giving mass to the up-quarks and Dirac neutrinos, while $v_d$ generates down-quark and charged lepton masses. The reality of $10_H$ implies that $v_u = v_d^{\ast} \equiv v_{10}$, and thus the ratio $r = |v_u/v_d| = 1$.  With $r=1$, the needed splitting between the top and bottom quark masses cannot be achieved.  Note that $r=1$  is a special case of the general $SO(10) \times U(1)_{PQ}$ models with $v_u \neq v_d^*$.
Such models have been studied, which find the phenomenological requirement $r \sim m_t/m_b$. A three generation analysis of fermion masses and mixings with a complex $10_H$ in Ref.  \cite{Joshipura:2011nn} shows that a realistic fit requires $r\sim  70$, which is well outside of the prediction of $r=1$ in the case of real $10_H$.  Thus we conclude that one $\overline{126}_H$ and one real $10_H$ is not realistic \cite{Bajc:2005zf}.

What about using one $\overline{126}_H$ and one $120_H$?  As shown in Ref. \cite{Bajc:2005zf}, this case also cannot reproduce fermion masses correctly.  The ratio $m_t/m_b$ comes out to be of order one, rather than the phenomenological value of $\sim 70$.  In addition, as we shall show, this model predicts the GUT scale mass ratio $m_\tau/m_b \simeq 3$, with any deviation of order $m_s/m_b \sim 5\%$.  As already noted, the ratio $m_\tau/m_b = (1.4 - 1.7)$ at the GUT scale in $SO(10)$ models under discussion.  Thus we conclude that only two Higgs fields being responsible for Yukawa couplings cannot be realistic.

When three Higgs fields are introduced, the choice of one $10_H$, one $120_H$ and one $\overline{126}_H$ appears attractive, as there in no replication here.  This choice can indeed lead to a realistic fermion mass spectrum, as we elaborate in the next section.  There would be two complex symmetric Yukawa coupling matrices in this case, along with one complex antisymmetric matrix.  If an alternative choice of one $\overline{126}_H$ and two copies of $120_H$ can lead to a realistic spectrum, that would have less parameters with one symmetric and two antisymmetric Yukawa matrices.  However, as we show in Appendix B, this choice would lead to the relation $m_\tau = 3 m_b$ with corrections of order 5\%, even when one allows for large off-diagonal contributions to the mass matrices from the $120_H$. Models with one $\overline{126}_H$ and two copies of $10_H$ would be realistic; however, these models would have three complex symmetric Yukawa matrices which have more parameters compared to the case of one $10_H$, one $120_H$ and one $\overline{126}_H$.  This completes the first part of the proof that $n_{10} = n_{120} = n_{126} = 1$ is the economic choice for the Yukawa sector.  To complete the proof we establish in the next section that this choice is indeed realistic.

%%%%%%%%%%%%%%%%%%%%%%%%%%%%%%%%%%%%%%%%%%%%%
%%%%%%%%%%%%%%%%%%%%%%%%%%%%%%%%%%%%%%%%%%%%%
\section{\boldmath{Realistic Fermion Spectrum with Minimal Yukawa Sector}}

As argued in the previous section, the minimal Yukawa sector of $SO(10)$ makes use of one real $10_H$, one real $120_H$ and one complex $\overline{126}_H$ of Higgs bosons that couple to the three families of fermions in the $16_F$ representation.  Here we proceed to establish the consistency of such a theory with observed fermion masses and mixings.

With no symmetry other than the gauge symmetry of $SO(10)$ imposed, the most general Yukawa interactions of the model can be written down as
\begin{equation}
\label{yukawa}
\mathcal{L}_{yuk}= 16_F (Y_{10}10_H+Y_{120}120_H+Y_{126}\overline{126}_H) 16_F.
\end{equation}
Here $Y_{10}$ and $Y_{126}$ are complex symmetric Yukawa matrices, while $Y_{120}$ is a complex antisymmetric matrix.
Under the Pati-Salam subgroup $G_{PS} \equiv SU(2)_L \times SU(2)_R \times SU(4)_c$, these fields decompose as
\begin{align}
&16=(2,1,4) + (1,2,\overline{4})\\
&10=(2,2,1)+(1,1,6)\\
&120=(2,2,1)+(1,1,10)+(1,1,\overline{10})+(3,1,6)+(1,3,6)+(2,2,15)\\
&126=(1,1,6)+(3,1,10)+(1,3,\overline{10})+(2,2,15).
\end{align}
The $10_H$ has one SM doublet Higgs field contained in the bidoublet (2,2,1), while the $120_H$ has two SM Higgs doublets, one each belonging to (2,2,1) and (2,2,15).  The reality condition for the (2,2,1) from $10_H$ is listed in Eq. (\ref{real}), while those from the $120_H$ would imply $v^{(1)}_u = v_d^{(1)\ast}\equiv v^{(1)}_{120}$ and $v^{(15)}_u = v_d^{(15)\ast}\equiv v^{(15)}_{120}$ with the superscripts $(1)$ and $(15)$ denoting the (2,2,1) and the (2,2,15) fragments.  The $126_H$ contains two SM Higgs fields contained in the complex bidoublet (2,2,15) fragment, which is not subject to the reality condition. We denote the up-type and down-type electroweak VEVs of the $\overline{126}_H$ as $v^u_{126}$ and $v^d_{126}$ respectively.  Note also that the $(1,3,10)$ fragment of $\overline{126}_H$ contains a SM singlet field which generates large Majorana masses for the right-handed neutrinos once it acquires a VEV.

The up-quark, down-quark, charged leptons, Dirac neutrino and Majorana neutrino mass matrices derived from Eq. (\ref{yukawa})can be now written down:
\vspace*{-0pt}
\begin{align}
&M_U= v_{10}Y_{10}+v^u_{126}Y_{126}+(v^{(1)}_{120}+v^{(15)}_{120})Y_{120},\\
&M_D= v^{\ast}_{10}Y_{10}+v^d_{126}Y_{126}+(v^{(1)\ast}_{120}+v^{(15)\ast}_{120})Y_{120},\\
&M_E= v^{\ast}_{10}Y_{10}-3v^d_{126}Y_{126}+(v^{(1)\ast}_{120}-3v^{(15)\ast}_{120})Y_{120},\\
&M_{\nu_D}= v_{10}Y_{10}-3v^u_{126}Y_{126}+(v^{(1)}_{120}-3v^{(15)}_{120})Y_{120},\\
&M_{\nu_{R,L}}=v_{R,L}Y_{126}. \label{MR}
\end{align}

\noindent Now defining
\vspace*{-0pt}
\begin{align}
D&= v_{10}Y_{10},\;\; A=(v^{(1)}_{120}+v^{(15)}_{120})Y_{120},\;\; S=v^u_{126}Y_{126},\\
r_1&=\frac{v^d_{126}•}{•v^u_{126}},\;\; r_2=\frac{•v^{(1)\ast}_{120}-3v^{(15)\ast}_{120}}{•v^{(1)}_{120}+v^{(15)}_{120}},\;\; e^{i\phi}= \frac{•v^{(1)\ast}_{120}+v^{(15)\ast}_{120}}{•v^{(1)}_{120}+v^{(15)}_{120}},\;\; c_{R,L}=\frac{•v_{R,L}}{•v^u_{126}},\label{cR}
\end{align}

\noindent and going into a phase convention where $v_{10}$ is real (this can be done by an $SU(2)_L$ rotation), we get
\vspace{-0pt}
\begin{align}
&M_{U} = D+S+A, \\
&M_{D} = D+r_1 S+ e^{i \phi} A, \\
&M_{E} = D-3 r_1 S+ r_2 A, \label{ME} \\
&M_{\nu_D} = D-3 S+ r^{\ast}_2 e^{i \phi} A, \label{MD}\\
&M_{\nu_{R,L}} = c_{R,L} S.
\end{align}

\noindent
These matrices are written in a basis $f_iM_{ij}f^c_j$.  The light neutrino mass matrix, obtained from the see-saw formula, is given by
\vspace{-5pt}
\begin{align}
\label{MN}
M_N = M_{\nu_L} - M_{\nu_D} M^{-1}_{\nu_R} M^T_{\nu_D} .
\end{align}

Without loss of generality one can choose a basis where $S$ is real, positive and diagonal.
In this basis, $S$ would have 3 real parameters while $D$ has 6 complex parameters. Since the matrix $A$ is antisymmetric, it has 3 complex parameters. There are 4 additional complex parameters in $r_{1,2}, c_{R,L}$ and one phase $\phi$. An overall phase either from $c_L$ or $c_R$  will be irrelevant in the matrix $M_N$. Altogether there are then 16 real parameters and 13 phases. With these parameters one should fit 18 observables:
6 quark masses, 3 quark mixing angles, 1 CKM phase, 3 charged lepton masses, 2 neutrino mass squared differences, and 3 mixing angles in the neutrino sector.  If we assume dominance of either type-I or type-II seesaw, then the parameter set is reduced by 1 magnitude and 1 phase.  Although the number of model parameters is larger than the number of observables, it is nontrivial to find an acceptable fit owing to the fact that 12 or 13 parameters are phases which cannot be manipulated much.

The type-II contribution to the light neutrino mass matrix originates in the model from terms such as $10^2_H 126^2_H$ in the scalar potential.
When decomposed into the the Pati-Salam symmetry group, this term would contain terms of the type $(3,1,10) (2,2,1)^2 (1,3,\overline{10})$.
When the singlet VEV of $(1,3,\overline{10})$ and the doublet VEV of $(2,2,1)$ are inserted in this term, a linear term in $(3,1,10)$ would result, which leads to an induced VEV for its neutral component: $v_L \sim v_R v^2/M^2_{\rm GUT}$.
We note that with the right-handed neutrino mass given as in \eqref{MR}, the mass of the $(X',Y')$ gauge bosons which are outside of $SU(5)$ but mediate proton decay is given as $M_{X',Y'}= \sqrt{2} g v_R$, where $g$ is the $SO(10)$ gauge coupling.

%%%%%%%%%%%%%%%%%%%%%%%%%%%%%%%
%%%%%%%%%%%%%%%%%%%%%%%%%%%%%%%
\subsection{Numerical analysis of the fermion masses and mixings}

In this section we discuss the procedure we follow for the numerical analysis to the fermion masses and mixings and present our fit results.  For optimization purpose we do
a $\chi^{2}$-analysis. The pull and $\chi^{2}$-function are defined as:
\vspace{-2pt}
\begin{align}
P_{i} &= \frac{O_{i\; \rm{th}}-E_{i\; \rm{exp}}}{\sigma_{i}}, \\
\chi^{2} &= \sum_{i} P_{i}^{2},
\end{align}

\noindent
where $\sigma_{i}$ represent experimental 1$\sigma$ uncertainty  and $O_{i\ \rm{th}}$, $E_{i\; \rm{exp}}$ and $P_{i}$ represent the theoretical prediction, experimental central value and pull of observable $i$. We fit the values of the observables at the GUT scale, $M_{GUT}=2\times 10^{16}$ GeV. To get the GUT scale values of the observables we take the central values at the $M_{Z}$ scale from Table-1 of Ref. \cite{Antusch:2013jca}. With this input we do the renormalization group equation (RGE) running of the Yukawa couplings \cite{Machacek:1983fi} and the CKM parameters \cite{Babu:1987im} within the SM up to the GUT scale.  For the associated one sigma uncertainties of the observables at the GUT scale, we keep the same percentage uncertainty with respect to the central value of each quantity as that at the $M_{Z}$ scale. For the charged lepton masses, a relative uncertainty of 0.1$\%$ is assumed in order to take into account the theoretical uncertainties arising for example from threshold effects.   All these inputs are presented in Table \ref{tab:04}.
The RGE running factors for the Yukawa couplings $\eta_i = y_i(M_{GUT})/y_i(M_Z)$ and for the CKM mixing angles
$\eta^{\rm{CKM}}_{ij} = \theta^{\rm{CKM}}_{ij}(M_{GUT})/\theta^{\rm{CKM}}_{ij}(M_Z)$ are taken to be:
\vspace*{-0pt}
\begin{align}
&(\eta_u, \eta_c, \eta_t)= (0.382, 0.382, 0.434) \label{rge-a} \\
&(\eta_d, \eta_s, \eta_b)= (0.399, 0.399, 0.348) \\
&(\eta_e, \eta_{\mu}, \eta_{\tau})= (0.967, 0.967, 0.967) \\
&(\eta^{\rm{CKM}}_{12}, \eta^{\rm{CKM}}_{23}, \eta^{\rm{CKM}}_{13})= (1.000, 1.154, 1.154) \label{rge-b}
\end{align}

\FloatBarrier
\begin{table}[th]
\centering
\footnotesize
\resizebox{0.7\textwidth}{!}{
\begin{tabular}{|c|c|c|}
\hline
\pbox{10cm}{Yukawa Couplings \\ \& CKM parameters} & $\mu= M_{Z}$  & $\mu= M_{GUT}$  \\ [1ex] \hline
$y_{u}/10^{-6}$ & $6.65 \pm 2.25$ &$2.54\pm0.86$ \\ \hline
$y_{c}/10^{-3}$ & $3.60 \pm 0.11$ &$1.37\pm0.04$  \\ \hline
$y_{t}$ & $0.9860 \pm 0.00865$ &$0.428\pm0.003$  \\ \hline
$y_{d}/10^{-5}$ & $1.645 \pm 0.165$ &$6.56\pm0.65$ \\ \hline
$y_{s}/10^{-4}$ & $3.125 \pm 0.165$ &$1.24\pm0.06$  \\ \hline
$y_{b}/10^{-2}$ & $1.639 \pm 0.015$ &$0.57\pm0.005$  \\ \hline
$y_{e}/10^{-6}$ & $2.79475 \pm 0.0000155$ &$2.70341\pm0.00270$  \\ \hline
$y_{\mu}/10^{-4}$ & $5.89986 \pm 0.0000185$ &$5.70705\pm0.00570$ \\ \hline
$y_{\tau}/10^{-2}$ & $1.00295 \pm 0.0000905$  &$0.97020\pm0.00097$ \\ \hline
$\theta^{\rm{CKM}}_{12}$ & $0.22735\pm 0.00072$ &$0.22739\pm0.0006$ \\ \hline
$\theta^{\rm{CKM}}_{23}/10^{-2}$ & $4.208\pm 0.064$ &$4.858\pm0.06$ \\ \hline
$\theta^{\rm{CKM}}_{13}/10^{-3}$ & $3.64\pm 0.13$ &$4.202\pm0.13$ \\ \hline
$\delta^{\rm{CKM}}$ & $1.208\pm0.054$ &$1.207\pm0.054$ \\ [0.5ex] \hline
\end{tabular}
}
\caption{  Values of observables at $M_{Z}$ scale from Ref. \cite{Antusch:2013jca}. Here  experimental central
values with associated 1$\sigma$ uncertainties are quoted. The masses of fermions are given by the relations $m_i = v \;y_i$
with $v=174.104$ GeV. The corresponding values at the GUT scale are obtained by RGE evolution. For the associated
one sigma uncertainties of the observables at the GUT scale, we keep the same percentage uncertainty with
respect to the central value of each quantity as that at the $M_{Z}$ scale.  For the charged lepton Yukawa
couplings at the GUT scale, a relative uncertainty of 0.1$\%$  is assumed in order to take into account the
theoretical uncertainties arising for example from threshold effects.}
\label{tab:04}
\end{table}

The low scale inputs as shown in Table \ref{nu}  in the neutrino sector are taken from Ref. \cite{Fogli:2012ua}.
For neutrino observables, we run the RGE for the dimension five operator from low scale to the $v_R$ scale
\cite{Babu:1993qv} and use these new values during the fitting produce. For this running purpose, we have assumed hierarchical structure of the neutrinos and used the approximations $m_2= \sqrt{•\Delta m^2_{sol}}$ and  $m_3= \sqrt{•\Delta m^2_{atm}}$.  The running values of the observables at the high scale depend on the scale
$v_R$, this is why we present the neutrino mass squared differences resulting from running in Table \ref{tab:01}
at the relevant scale $v_R$ corresponding to two different fits (type-I dominance and type-I+II case), while all the other inputs are at $M_{\rm GUT}=2\times10^{16}$ GeV.

In $SO(10)$ GUT models such as the one we are considering, the (3,3) entry of the Dirac neutrino Yukawa coupling matrix $Y_{\nu D}$ is expected to be of the order of unity, and thus RGE corrections proportional to $Y_{\nu D}$ can be important in the momentum range $M_{\nu^c} \leq \mu \leq M_{\rm GUT}$. This effect
could have a sizeable contribution to the tau lepton mass only, since for the first and second generation Dirac Yukawa couplings turn out to be small.
Including this effect of the heavy right-handed neutrinos thresholds,
the Dirac neutrino mass matrix gets modified at the GUT scale as
\vspace{-5pt}
\begin{align}\label{MDp}
M^{\prime}_{\nu_D} = \left[1-\frac{3}{2(16 \pi^2)} Y_{\nu D} \log(\frac{M_{GUT}}{c_R S}) Y^{\dagger}_{\nu D}\right]  M_{\nu_D}.
\end{align}

\noindent  while the modified charged lepton mass matrix becomes
\vspace{-5pt}
\begin{align}\label{MEp}
M^{\prime}_E = \left[1+\frac{3}{2(16 \pi^2)} Y^{\prime}_{\nu D} \log(\frac{M_{GUT}}{c_R S}) Y^{\prime \dagger}_{\nu D}\right] M_{E} .
\end{align}

\noindent To be clear, the tau lepton mass decreases in going from the $\nu_R$ mass scale to the GUT scale due to the Dirac neutrino Yukawa correction. In the fitting procedure it was thus $M_E^\prime$ from Eq. \eqref{MEp} to be compared to the
experimental values at $M_{GUT}$, while in \eqref{MN} $M_{\nu D}$ has been replaced by $M_{\nu D}^\prime$ \eqref{MDp}. Notice that
$M_E$ and $Y_{\nu D}=M_{\nu D}/v$ in Eqs. \eqref{MEp} and  \eqref{MDp} are defined in Eqs. \eqref{ME} and  \eqref{MD}.

We investigate three different scenarios, type-I dominance, type-II dominance and the general scenario where both contributions are present, type-I+II. The fit results corresponding to our numerical analysis is  presented in Table \ref{tab:01}. We found good solutions for both type-I and type-I+II with total $\chi^{2}=$ 0.45 and 0.004 respectively, but not for type-II scenario (the total $\chi^2 \sim 1000$ in this case).  For the type-II case, our numerical analysis shows that, for the best fit, the worst fitted quantity  corresponds to $\Delta m^2_{sol}$ that comes out to be $\sim 10^3$ times smaller (with pull $\sim$ -32) than the experimental data.  The other  discrepancy  is of the quantity $\theta^{PMNS}_{23}$ that is $\sim$1.5 times smaller compared to the experimental central value. With these fit results the predictions of the model for these two scenarios are listed in Table \ref{tab:02}. The parameter set corresponding to these best fit results are presented in Appendices C and D for type-I and type-I+II cases respectively.  We conclude that the model gives an excellent fit to all observables in the fermion sector.  This completes our proof of the minimality of the Yukawa sector in $SO(10)$ models.

\begin{table}[th]
\centering
\begin{tabular}{|c|c|}
\hline
Quantity & Central Value \\ \hline
$\Delta m^{2}_{sol}/10^{-5} eV^{2}$ &7.56$\pm$0.24  \\ \hline
$\Delta m^{2}_{atm}/10^{-3} eV^{2}$ &2.41$\pm$0.08 \\ \hline
$\sin^{2}\theta^{\rm{PMNS}}_{12}/10^{-1}$  &3.08$\pm$0.17 \\ \hline
$\sin^{2}\theta^{\rm{PMNS}}_{23}/10^{-1}$ &3.875$\pm$0.225 \\ \hline
$\sin^{2}\theta^{\rm{PMNS}}_{13}/10^{-2}$ &2.41$\pm$0.25   \\ [1ex] \hline
\end{tabular}
\caption{  Observables in the neutrino sector used in our fits taken from Ref. \cite{Fogli:2012ua}. }\label{nu}
\end{table}

\FloatBarrier
\begin{table}[th]
\centering
\footnotesize
\resizebox{1.0\textwidth}{!}{
\begin{tabular}{|c|c||c|c||c|c||}
\hline
\pbox{10cm}{Masses (in GeV) and \\  Mixing parameters} & \pbox{10cm}{~~~~~Inputs \\ (at $\mu= M_{GUT}$)} & \pbox{25cm}{Fitted values \\ (\textbf{type-I})\\ (at $\mu= M_{GUT}$)} & \pbox{10cm}{~~~pulls  \\ (\textbf{type-I})} & \pbox{25cm}{Fitted values \\ (\textbf{type-I+II})\\ (at $\mu= M_{GUT}$)  } & \pbox{10cm}{~~~~pulls  \\ (\textbf{type-I+II})}  \\ [1ex] \hline
$m_{u}/10^{-3}$ & 0.442$\pm$0.149 & 0.444 & 0.009 &0.442 &-0.0002 \\ \hline
$m_{c}$   & 0.238$\pm$0.007 & 0.238 & -0.002&0.238 &0.0001  \\ \hline
$m_{t}$   & 74.51$\pm$0.65 & 74.52 & 0.009 &74.52 &-0.005 \\ \hline
$m_{d}/10^{-3}$  & 1.14$\pm$0.11 & 1.14 & -0.0002 &1.14 &-0.00006 \\ \hline
$m_{s}/10^{-3}$  & 21.58$\pm$1.14 & 21.60 & 0.007 &21.59 &0.0001  \\ \hline
$m_{b}$  & 0.994$\pm$0.009 & 0.994 & 0.002 &0.994 &0.000005 \\ \hline
$m_{e}/10^{-3}$   & 0.470692$\pm$0.000470 & 0.470674 & -0.03 &0.470675 &-0.003 \\ \hline
$m_{\mu}/10^{-3}$  & 99.3658$\pm$0.0993 & 99.3618 & -0.04 &99.3621 &-0.003 \\ \hline
$m_{\tau}$  & 1.68923$\pm$0.00168 & 1.68925 & 0.01 &1.68925 &0.001  \\ \hline
$|V_{us}|/10^{-2}$  & 22.54$\pm$0.06 & 22.54 & 0.002 &22.54 &0.00008 \\ \hline
$|V_{cb}|/10^{-2}$  & 4.856$\pm$0.06 & 4.856 & 0.001 &4.856 &0.0007 \\ \hline
$|V_{ub}|/10^{-2}$  & 0.420$\pm$0.013 & 0.420 & -0.007 &0.420 &-0.0001 \\ \hline
$\delta_{CKM}$ & 1.207$\pm$0.054  & 1.207 & 0.01 &1.207 &0.005 \\ \hline
$\Delta m^{2}_{sol}/10^{-4}$(eV$^{2}$) & \pbox{10cm}{1.29$\pm$0.04 ($1\times10^{15}$GeV)\\1.27$\pm$0.04 ($7.3\times10^{12}$GeV)} & 1.27 & -0.48 &1.27 &0.04 \\ \hline
$\Delta m^{2}_{atm}/10^{-3}$(eV$^{2}$) & \pbox{10cm}{4.12$\pm$0.13 ($1\times10^{15}$GeV)\\4.05$\pm$0.13 ($7.3\times10^{12}$GeV)} & 4.06 &- 0.46 &4.06 &0.04\\ \hline
$\sin^{2}\theta^{\rm{PMNS}}_{12}$ & 0.308$\pm$0.017 & 0.308 & -0.01 &0.308 &0.00001  \\ \hline
$\sin^{2}\theta^{\rm{PMNS}}_{23}$ & 0.387$\pm$0.0225 & 0.387 & -0.01 &0.387 &-0.00006  \\ \hline
$\sin^{2}\theta^{\rm{PMNS}}_{13}$ & 0.0241$\pm$0.0025 & 0.0241 & 0.01 & 0.0241&-0.0003 \\ \hline
\end{tabular}
}
\caption{ Best fit values of the observables correspond to $\chi^{2}=$ 0.45 and 0.004 for \textbf{type-I} and \textbf{type-I+II} scenarios respectively for 18 observables. For the charged lepton masses, a relative uncertainty of 0.1$\%$  is assumed in order to take into account the theoretical uncertainties arising for example from threshold effects. The neutrino mass squared differences are fitted at the $v_R$ scale, which for our solutions are $\sim 1\times 10^{15}$ GeV and $\sim 7.3\times 10^{12}$ GeV for \textbf{type-I} and \textbf{type-I+II} respectively. Here the $v_R$ scale is determined by using the relation $v_R= c_R v^u_{126}$ given in Eq. \eqref{cR}, we have taken $v^u_{126}=174.104$ GeV.  One should note that due to the right-handed neutrino threshold corrections the charged lepton mass matrix gets modified and is given in Eq. \eqref{MEp}. The fitted masses for the charged leptons presented in this table are the eigenvalues of this modified matrix, $M^{\prime}_E$. The effect of the right-handed neutrinos is to  decrease the tau lepton mass in going from $\nu_R$ scale to the  GUT scale. For the fits presented in the table, the actual fitted  mass of the tau lepton is $m_{\tau}=1.617$ GeV  ($1.573$ GeV) at the GUT scale for the   \textbf{type-I} (\textbf{type-I+II}) scenario, which matches correctly with the input value when the right-handed neutrino threshold correction is taken into account.  For \textbf{type-II} scenario, we have not found any acceptable solution as mentioned in the text.    }
\label{tab:01}
\end{table}

\FloatBarrier
\begin{table}[th]
\centering
\footnotesize
\resizebox{1.0\textwidth}{!}{
\begin{tabular}{|c|c|c|}
\hline
Quantity & \pbox{10cm}{Predicted Value  (\textbf{type-I})} & \pbox{10cm}{Predicted Value  (\textbf{type-I+II})}   \\ [1ex] \hline
$\{m_{1}, m_{2}, m_{3} \}$ (in eV) & $\{ 1.51\cdot10^{-4}, 1.12\cdot10^{-2}, 6.47\cdot10^{-2} \}$  & $\{ 1.02\cdot10^{-2}, 1.52\cdot10^{-2}, 6.55\cdot10^{-2} \}$   \\ \hline

$\{\delta^{PMNS}, \alpha^{PMNS}_{21}, \alpha^{PMNS}_{31} \}$ & $\{ 2.81^{\circ}, 169.61^{\circ},27.25^{\circ} \}$ & $\{ -150.82^{\circ}, -136.92^{\circ}, -106.50^{\circ} \}$ \\ \hline

$\{m_{cos}, m_{\beta}, m_{\beta \beta} \}$ (in eV) & $\{ 7.61\cdot10^{-2}, 5.05\cdot10^{-3}, 2.13\cdot10^{-3} \}$  & $\{ 9.10\cdot10^{-2},1.30\cdot10^{-2}, 4.09\cdot10^{-3} \}$ \\ \hline

$\{M_{1}, M_{2}, M_{3} \}$ (in GeV)  & $\{ 1.04\cdot10^{5}, 1.23\cdot10^{12}, 4.34\cdot10^{14} \}$ & $\{ 6.14\cdot10^{6}, 1.12\cdot10^{10}, 3.14\cdot10^{12} \}$  \\ \hline
\end{tabular}
}
\caption{ Predictions of the minimal non-SUSY $SO(10)$ model for \textbf{type-I}  and \textbf{type-I+II}  scenarios. $m_{i}$ are the light neutrino masses, $M_{i}$ are the right handed neutrino masses, $\alpha_{21,31}$ are the Majorana phases following the PDG parametrization, $m_{cos}=\sum_{i} m_{i}$, $m_{\beta}=\sum_{i} |U_{e i}|^{2} m_{i}$ is the effective mass parameter for beta-decay and $m_{\beta \beta}= | \sum_{i} U_{e i}^{2} m_{i} |$ is the effective mass parameter for neutrinoless double beta decay.}
\label{tab:02}
\end{table}

%%%%%%%%%%%%%%%%%%%%%%%%%%%%%
%%%%%%%%%%%%%%%%%%%%%%%%%%%%%
\section{Gauge Coupling Unification}

As is well known, the three gauge couplings of the SM do not unify at a common scale.  $SO(10)$ models provide a way to achieve coupling unification by virtue of an intermediate scale.  In our proposed framework, the first stage of symmetry breaking can be achieved by employing a real $45_H$, or a real $54_H$ or a real $210_H$, along with a complex $126_H$.  Employing $45_H$ Higgs would require relying on the quantum corrections in the Higgs potential ~\cite{Bertolini:2009es, Bertolini:2010ng,Graf:2016znk}, while there is no such problem with the use of $210_H$.  In both cases the discrete D Parity symmetry would be broken at the GUT scale \cite{Chang:1983fu}. The intermediate gauge symmetry may be $SU(2)_L \times SU(2)_R \times SU(4)_c$ when a $210_H$ is used, while it is $SU(2)_L \times SU(2)_R \times SU(3)_c \times U(1)_{B-L}$ if the $45_H$ is used.  Alternatively, a $54_H$ can break $SO(10)$ down to Pati-Salam symmetry preserving D parity.  In this case the unification scale tends to be lower, of order $2 \times 10^{15}$ GeV, if threshold effects arising from the scalar multiplets are ignored. This can potentially be in conflict with proton decay limits.  It has been recently shown in Ref. \cite{khan} that symmetry breaking with a $54_H$ and a $126_H$ can lead to higher values of $M_{\rm GUT}$ consistent with proton lifetime, when threshold effects are properly included. Here we present for completeness our results on the unification of gauge couplings assuming the intermediate symmetry to be $SU(2)_L \times SU(2)_R \times SU(4)_c$ with or without D parity.

\begin{figure}[t]
\centering
\includegraphics[scale=0.72]{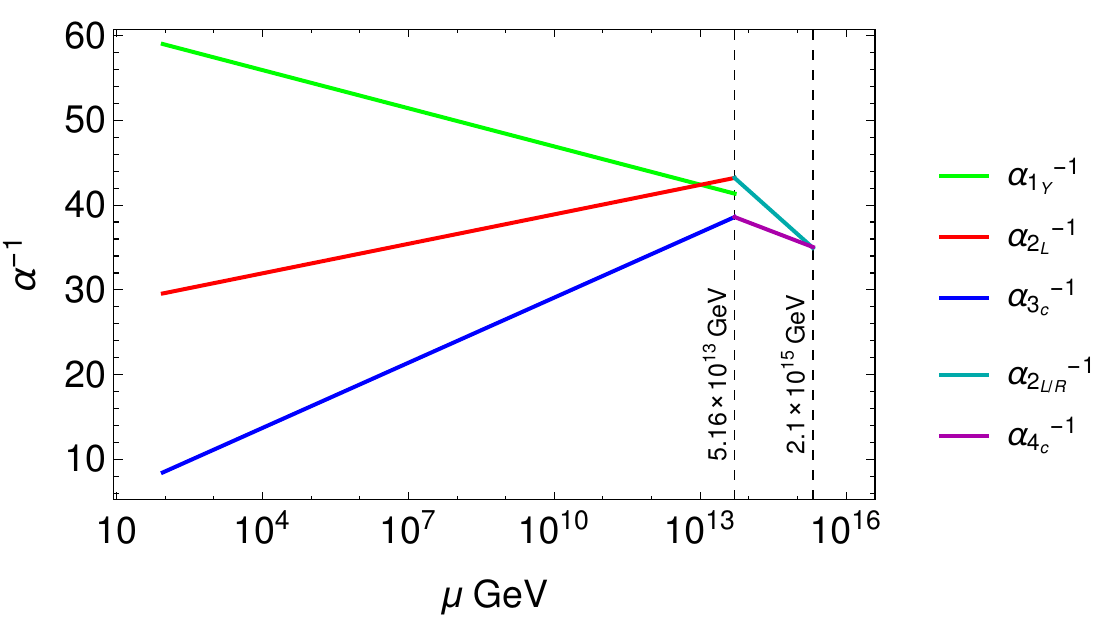}
\includegraphics[scale=0.72]{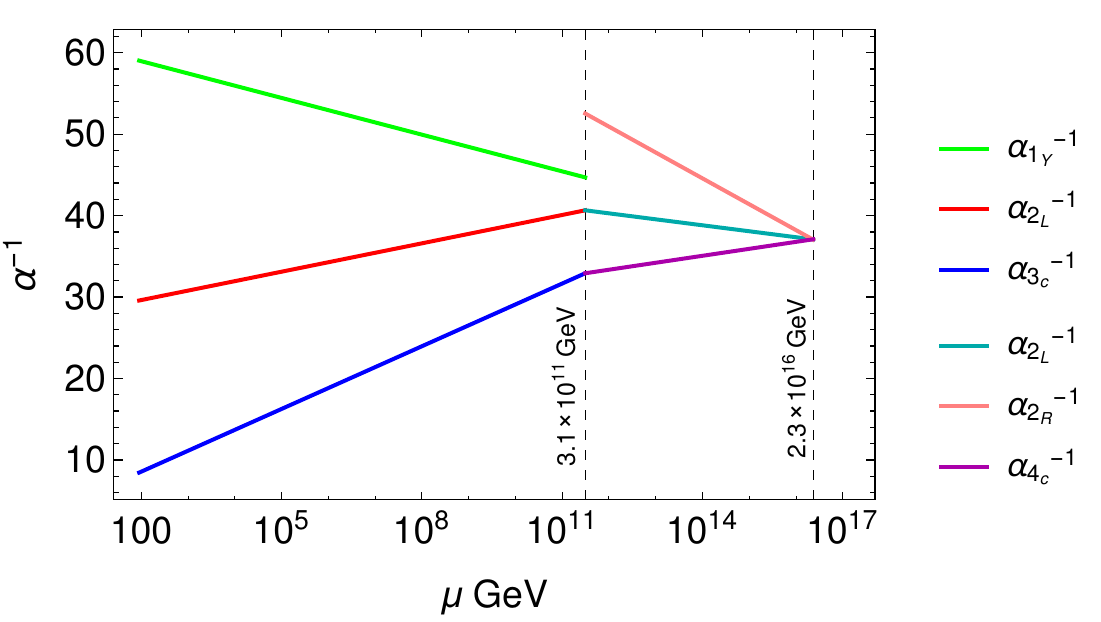}
\caption{1-loop gauge coupling running of the three SM gauge couplings from low scale to intermediate PS scale and from PS scale to GUT scale for minimal non-SUSY $SO(10)$ model. The left plot corresponds to the case when the GUT symmetry is broken by $54_H$ Higgs that leaves the discrete symmetry $g_L=g_R$ unbroken. The right plot is for the case when $54_H$ is replaced by  $210_H$  Higgs that does not preserve the discrete symmetry. }\label{uni0}
\end{figure}

Since breaking $SO(10)$ gauge group by $54_H$ Higgs preserves the discrete parity,
that demands the equality of the $SU(2)_L$ and $SU(2)_R$ gauge couplings ($g_L=g_R$) at the PS intermediate scale. The low energy data completely determines the value of this scale  as well as the GUT scale with the assumption of survival hypothesis
\cite{Georgi:1979md,delAguila:1980qag,Mohapatra:1982aq}.  The one-loop beta function coefficients for the evolution of the $U(1)$, $SU(2)_L$ and $SU(3)_c$ gauge couplings are $b_i=\{ 41/10, -19/6, -7 \}$ \cite{Jones:1981we}. To determine the intermediate scale, we use the low energy values from~ Ref.\cite{Antusch:2013jca}: $\alpha^{-1}_{1}(M_{Z})=59.02$, $\alpha^{-1}_{2}(M_{Z})=29.57$ and $\alpha^{-1}_{3}(M_{Z})=8.44$  (only the central values are quoted here). Then from the intermediate scale to the GUT scale we run the RGEs with one-loop coefficients $b_i=\{ 67/6, 67/6, 10/3 \}$ for the group $G_{224}$ that determines the GUT scale. The existence of the multiplets $(2,2,1)\subset 10_H$, $(2,2,1)+ (2,2,15)\subset 120_H$ and $(2,2,15)+ (3,1,10)+ (1,3,\overline{10})\subset 126_H$ is assumed at the intermediate scale while the rest of the multiplets are assumed to have GUT scale mass following the survival hypothesis. One-loop running  of the RGEs of the gauge couplings are shown in Fig. \ref{uni0} (left plot).

%\FloatBarrier

From this Fig.  \ref{uni0} one sees that the GUT scale is $\sim 2\times 10^{15}$ GeV, which is about a factor of 2.5
smaller  compared to what is needed to save the theory from the experimental proton decay limit $\tau_p \gtrsim 1.29
\times 10^{34}$ yrs  \cite{Nishino:2012bnw}. Certainly the assumption made that all scalar particles have a common mass at the assumed scale is too
restrictive: the Higgs multiplets are likely to have non-degenerate mass spectrum with masses scattered
around each scale under consideration. If one includes this threshold correction arising from the Higgses, the unification
scale can be raised as shown in an explicit calculation in Ref. \cite{khan}. There is no
strict guideline, however, on how much the mass spectrum may be scattered; this would lead to significant uncertainty in proton lifetime estimate.
As we show in the next section, the branching ratios for proton decay are much more stable and can be used to test these theories.

If instead of a $54_H$ a $210_H$ is used to break the GUT symmetry, then the unification scale is naturally raised to about $2 \times 10^{16}$ GeV.  This is because D parity is broken by the VEV of $210_H$, and as a result, with the assumption of survival hypothesis, the intermediate scale scalar spectrum is left-right asymmetric.  Although $210_H$ allows for other intermediate symmetries, here we focus on the Pati-Salam  symmetry. The gauge coupling evolution with the PS intermediate symmetry is presented in Fig. \ref{uni0} (right plot). Following survival hypothesis, we consider  the multiplets $(2,2,1)\subset 10_H$ and $(2,2,15)+  (1,3,\overline{10})\subset 126_H$ at the intermediate scale with the rest of the multiplets lying at the GUT scale. With these multiplets, the one-loop RGE coefficients are $b_i=\{2, 26/3, -7/3\}$ for the group $G_{224}$. This plot clearly shows that the GUT scale can be raised by about an order of magnitude compared to the $54_H$ scenario and one does not need to rely on the threshold correction to save the theory from rapid proton decay. It should be noted that the scenario with $210_H$ has a drawback that the intermediate scale is relatively low $\sim 10^{11}$ GeV, which does not fit the right-handed neutrino mass spectrum as well as the $54_H$ model where this scale is around $(10^{13} - 10^{14})$ GeV.  A look at the heaviest right-handed neutrino mass from Table \ref{tab:02} suggests that the case of type-I seesaw prefers symmetry breaking by a $54_H$, while the type I + type II scenario can accommodate breaking by a $210_H$.

%%%%%%%%%%%%%%%%%%%%%%%%%%%%%%%%%%%%%%%
%%%%%%%%%%%%%%%%%%%%%%%%%%%%%%%%%%%%%%%
\section{Proton Decay Branching Ratios} \label{D}

In non-SUSY $SO(10)$ models, proton decay mediated by the gauge bosons are the most important.   The lifetime of the proton is extremely sensitive to the superheavy gauge bosons masses ($M_{{(X,Y)}}$) since the lifetime goes as $\tau_p \sim M^4_{(X,Y)}/(g^4 m^5_p)$, where $m_p$ is the proton mass and $g$ is the unified gauge coupling. As noted in the previous section, there is a large uncertainty in the determination of $M_X$ from low energy data, owing to unknown high scale threshold effects.  On the other hand, proton decay branching ratios are less sensitive to these threshold effects, and so we focus on the predictions of the model for branching ratios.

The gauge bosons of $SO(10)$ belong to the adjoint $45$. The decomposition of this field under the SM gauge symmetry is given by:
\vspace*{0pt}
\begin{align}
45&=(1,1,1)+(1,1,0)+(1,1,-1)+(1,3,0)+(3,2,\frac{1}{6})+(3,2,-\frac{5}{6})+(\overline{3},2,\frac{1}{6})+(\overline{3},2,\frac{5}{6})
\nonumber \\
&+(1,1,0)+(3,1,\frac{2}{3})+(\overline{3},1,-\frac{2}{3})+(8,1,0).
\end{align}

\noindent The gauge bosons responsible for proton decay are the $(X,Y) (3,2,-5/6)$ and $(X^{\prime},Y^{\prime}) (3,2,1/6)$.  The gauge interaction Lagrangian of these bosons with the fermions in the current eigenstate basis is given  by \cite{Machacek:1979tx}:
\vspace*{0pt}
\begin{align}
\mathcal{L}_{int} &= \frac{g}{\sqrt{2}} \{
[-\overline{e}_L\gamma^{\mu}\overline{X}^{i}_{\mu}d^C_{Li}]
+
[\overline{\nu}_L\gamma^{\mu}\overline{Y}^{i}_{\mu}d^C_{Li}]
+
[\overline{d}_{L i}\gamma^{\mu}\overline{X}^{i}_{\mu}e^C_{L}
+
\epsilon_{ijk}\overline{u}^{Ck}_L\gamma^{\mu}\overline{X}^{i}_{\mu}
u^{j}_L]
\nonumber \\ &
+
[-\overline{u}_{L i}\gamma^{\mu}\overline{Y}^{i}_{\mu}e^C_{L}
+
\epsilon_{ijk}\overline{u}^{Ck}_L\gamma^{\mu}\overline{Y}^{i}_{\mu}
d^{j}_L]
+
[-\epsilon_{ijk}\overline{d}^{Ck}_L\gamma^{\mu}\overline{X}^{\prime i}_{\mu}
d^{j}_L]
+
[-\overline{u}_{R i}\gamma^{\mu}\overline{X}^{\prime i}_{\mu}
\nu^C_R]
+
[-\overline{u}_{L i}\gamma^{\mu}\overline{X}^{\prime i}_{\mu}
\nu^C_L]
\nonumber \\ &
+
[\epsilon_{ijk}\overline{d}^{C k}_L\gamma^{\mu}\overline{Y}^{\prime i}_{\mu}
u^{j}_L]
+
[-\overline{u}_{R i}\gamma^{\mu}\overline{Y}^{\prime i}_{\mu}
e^+_R]
+
[-\overline{d}_{L i}\gamma^{\mu}\overline{Y}^{\prime i}_{\mu}
\nu^C_L]
+ h.c.
\},
\end{align}

\noindent where $i,j,k$ are color indices and we have suppressed the family indices and $SU(2)_L$ indices.

The resulting $d=6$ effective operators of the form $QQQL$ responsible for proton decay can be constructed from this Lagrangian \cite{Nath:2006ut}:
\begin{align}
\mathcal{O}^{B-L}_I&=k_1^2 \; \epsilon_{ijk}\; \epsilon_{\alpha\beta}\; \overline{u^C_{ia}}_L\; \gamma^ \mu\; Q_{j\alpha aL}\; \overline{e^C_{b}}_L \;\gamma_\mu \;Q_{k \beta bL};\\
\mathcal{O}^{B-L}_{II}&=k_1^2 \; \epsilon_{ijk}\; \epsilon_{\alpha\beta}\; \overline{u^C_{ia}}_L\; \gamma^ \mu\; Q_{j\alpha aL}\; \overline{d^C_{kb}}_L \;\gamma_\mu \;L_{\beta bL};\\
\mathcal{O}^{B-L}_{III}&=k_2^2 \; \epsilon_{ijk}\; \epsilon_{\alpha\beta}\; \overline{d^C_{ia}}_L\; \gamma^ \mu\; Q_{j\beta aL}\; \overline{u^C_{kb}}_L \;\gamma_\mu \;L_{\alpha bL}; \\
\mathcal{O}^{B-L}_{IV}&=k_2^2 \; \epsilon_{ijk}\; \epsilon_{\alpha\beta}\; \overline{d^C_{ia}}_L\; \gamma^ \mu\; Q_{j\beta aL}\; \overline{\nu^C_{b}}_L \;\gamma_\mu \;Q_{k \alpha bL}.
\end{align}

\noindent Here, $k_1=g_{u}/(\sqrt{2}M_{(X,Y)})$ and $k_2=g_{u}/(\sqrt{2}M_{(X',Y')}$, $Q_L=(u_L,d_L)$ and $L_L=(\nu_L,e_L)$. The indices $i,j,k$ are color indices, $a,b$ are family indices and $\alpha,\beta$ are $SU(2)_L$ indices. In the physical basis these operators will be modified as:
\begin{align}
&\mathcal{O}(e_\alpha^C,d_\beta)= c(e_\alpha^C,d_\beta)\; \epsilon_{ijk}\; \overline{u^C_i}_L\; \gamma^\mu u_{jL} \;\overline{e^C_\alpha}_L\; \gamma_\mu \; d_{k\beta L};  \\
&\mathcal{O}(e_\alpha,d_\beta^C)= c(e_\alpha,d_\beta^C)\; \epsilon_{ijk}\; \overline{u^C_i}_L\; \gamma^\mu u_{jL}\; \overline{d^C_{k \beta}}_L\; \gamma_\mu \;e_{\alpha L};  \\
&\mathcal{O}(\nu_l, d_\alpha,d_\beta^C)= c(\nu_l, d_\alpha,d_\beta^C)\; \epsilon_{ijk}\; \overline{u^C_i}_L\; \gamma^\mu\; d_{j\alpha L}\; \overline{d^C_{k \beta}}_L\; \gamma_\mu\; \nu_{lL}; \\
&\mathcal{O}(\nu_l^C, d_\alpha,d_\beta^C)= c(\nu_l^C, d_\alpha,d_\beta^C)\; \epsilon_{ijk}\; \overline{d^C_{i \beta}}_L \; \gamma^\mu\; u_{j L}\; \overline{\nu_l^C}_L \;\gamma_\mu \;d_{k \alpha L};
\end{align}

\noindent where
\begin{align}
c(e^C_\alpha,d_\beta) &= k_1^2 \left[ V_1^{11} V_2^{\alpha \beta} +\left( V_1 V_{UD}\right)^{1\beta}\left(V_2 V^\dagger_{UD}\right)^{\alpha 1} \right]; \label{CIA} \\
c(e_\alpha,d^C_\beta) &= k_1^2 V_1^{11} V_3^{\beta \alpha}+k_2^2 \left(V_4V^\dagger_{UD}\right)^{\beta 1} \left( V_1 V_{UD} V_4^\dagger V_3\right)^{1 \alpha};  \\
c(\nu_l,d_\alpha,d_\beta^C)&=k_1^2 \left( V_1 V_{UD}\right)^{1\alpha}\left( V_3 V_{EN}\right)^{\beta l} +k_2^2 V_4^{\beta \alpha} \left( V_1 V_{UD} V_4^\dagger V_3 V_{EN}\right)^{1l};\\
c(\nu_l^C,d_\alpha,d_\beta^C)&= k_2^2 \left[\left(V_4 V_{UD}^\dagger \right)^{\beta 1} \left(U_{EN}^\dagger V_2\right)^{l \alpha} + V_4^{\beta \alpha} \left(U_{EN}^\dagger V_2 V_{UD}^\dagger\right)^{l 1}\right]; \alpha=\beta \neq 2. \label{CIB}
\end{align}

\noindent In the above $V_1, V_2$ etc are mixing matrices defined so that
 $V_1= U_L^T U_R$, $V_2=E_L^TD_R$,
$V_3=D_L^TE_R$, $V_4=D_L^T D_R$, $V_{UD}=U_R^{\dagger}D_R$,
$V_{EN}=E_R^{\dagger}N$ and $U_{EN}= E^T_L N$,
where $U,D,E$ define the  diagonalizing matrices given by
\vspace{-2pt}
\begin{eqnarray}
U^{\dagger}_L \ M_U \ U_R &=& M_U^{diag} \\
D^{\dagger}_L \ M_D \ D_R &=& M_D^{diag} \\
E^{\dagger}_L \ M_E \ E_R &=& M_E^{diag} \\
N^T \ M_N \ N &=& M_N^{diag} .
\end{eqnarray}

Then the partial decay width of the decay $N\to P+ \overline{l}$ ($N=p, n$, $P= (\pi, K, \eta)$ and $\overline{l}$ is anti-lepton) is given by:
\vspace*{-10pt}
\begin{align}
\Gamma(N\to P+ \overline{l})= \frac{m_N}{32 \pi} [1-(\frac{m_P}{m_N})^2]^2 \left| \sum_I C^I W^I_0(N\to P) \right|^2 ,
\end{align}

\noindent where the coefficients $C^I$ are given in Eqs. \eqref{CIA}-\eqref{CIB} and the relevant form factors $W_0$ are obtained by using lattice QCD computations  \cite{Aoki:2013yxa}:
\vspace*{-5pt}
\begin{align}
&\langle \pi^0|(u,d)_R u_L|p\rangle_0= -0.103, \langle \pi^+|(u,d)_R u_L|p\rangle_0= -0.103, \langle K^0|(u,s)_R u_L|p\rangle_0= 0.098, \nonumber \\
& \langle K^+|(u,s)_R d_L|p\rangle_0= -0.054, \langle K^+|(u,d)_R s_L|p\rangle_0= -0.093, \langle \eta|(u,d)_R u_L|p\rangle_0= 0.015. \nonumber
\end{align}

In Table \ref{tab:03} we present the $d=6$ proton decay   branching ratios calculated  for our best fit parameter sets. We find that the two dominant modes are $p\rightarrow  \overline{\nu} \pi^+$ and $p\rightarrow e^+\pi^0$. A comparison of these modes with those of more general $d=6$ proton decay studies \cite{Nath:2006ut} shows similarity.  The near dominance of the $\overline{\nu} \pi^+$ mode may be taken as a test of the Yukawa sector presented here.

\FloatBarrier
\begin{table}[th]
\centering
\footnotesize
\resizebox{0.4\textwidth}{!}{
\begin{tabular}{|c|c|c|}
\hline
p decay modes & \textbf{type-I} & \textbf{type-I+II}  \\ [1ex] \hline
$p\rightarrow \overline{\nu} + \pi^+ $ &49.07$\%$ &48.77$\%$ \\ \hline
$p\rightarrow e^+ \pi^0 $ &42.57$\%$  &35.16$\%$  \\ \hline
$p\rightarrow \mu^+ K^0 $ &4.13$\%$ &5.12$\%$ \\ \hline
$p\rightarrow \mu^+ \pi^0 $ &1.60$\%$ &5.62$\%$ \\ \hline
$p\rightarrow \overline{\nu} K^+ $ &1.19$\%$ &2.64$\%$ \\ \hline
$p\rightarrow e^+ K^0$ &0.99$\%$ &2.28$\%$ \\ \hline
$p\rightarrow e^+ \eta$  &0.40$\%$ &0.33$\%$ \\ \hline
$p\rightarrow \mu^+ \eta$  &0.01$\%$ &0.05$\%$ \\ [0.5ex] \hline
\end{tabular}
}
\caption{Proton decay branching ratios in minimal non-SUSY $SO(10)$ GUT in \textbf{type-I} and \textbf{type-I+II} cases.  For neutrino final states, we sum over all three flavors.  }
\label{tab:03}
\end{table}

%%%%%%%%%%%%%%%%%%%%%%%%%%%%%%%%%%%%%%%
%%%%%%%%%%%%%%%%%%%%%%%%%%%%%%%%%%%%%%%
\section{Conclusion}

In this work, we have presented an economic Yukawa sector for $SO(10)$ models.  The main feature of this construction is that only the $SO(10)$ symmetry is used to constrain the Yukawa parameters. The Higgs system consists of a {\it real} $10_H$, a {\it real} $120_H$ and a complex $\overline{126}_H$ that have Yukawa couplings.  In most nonsupersymmetric $SO(10)$ models in the literature symmetries outside of $SO(10)$ -- such as a Peccei-Quinn $U(1)$ -- are used to constrain the Yukawa sector.  That would require the complexification of the {\it real} $10_H$ and {\it real} $120_H$.  The model presented here deviates from this, and yet is quite constraining.  We showed that, with a limited number of Yukawa parameters, a good fit to
 all fermion masses and mixings, including the neutrino sector is possible.  Once the flavor sector is fixed, we are able to calculate the proton decay branching ratios.  The dominant decays of the proton are found to be  $p\rightarrow \overline{\nu} \pi^+$ and $p\rightarrow e^+ \pi^0$, which may provide partial tests of the model.

%%%%%%%%%%%%%%%%%%%%%%%%%%%%%%%%%%%%%%%%%%%%%
%%%%%%%%%%%%%%%%%%%%%%%%%%%%%%%%%%%%%%%%%%%%%
\section*{Acknowledgments}
This work is supported in part by the U.S. Department of Energy Grant No. de-sc0010108 (K.S.B and S.S)
and by the Slovenian Research Agency (B.B.). The authors wish to thank Barbara Szczerbinska and the CETUP* 2016 workshop for hospitality and for providing a stimulating environment for discussions. Numerical calculations are performed  with Cowboy Supercomputer of the High Performance Computing Center at Oklahoma State University (NSF grant no. OCI-1126330).

%%%%%%%%%%%%%%%%%%%%%%%%%%%%%%%%%%%%%%%
%%%%%%%%%%%%%%%%%%%%%%%%%%%%%%%%%%%%%%%
\begin{appendices}
%\appendixpageoff

\section{A comment on doublet-triplet splitting}

As is well known, any grand unified theory has to address the question of making one Higgs doublet light, while its color triplet GUT partner remains superheavy so as to not cause rapid proton decay.  This doublet-triplet splitting problem is present in both SUSY and
non-SUSY minimal GUTs. If the Higgs doublet mass is not split from the color triplet mass, either the electroweak symmetry would break at the GUT scale, or not break at all, or the light color-triplet would lead to far too fast proton decay.  A fine-tuning is necessary to bring the Higgs doublet mass down to the weak scale.  In supersymmetric versions, this fine-tuning is done at the tree level, SUSY would guarantee its stability against quantum corrections.  In non-supersymmetric $SO(10)$ theories, the tuning must be done after taking account of loop corrections to a very high order. The induced Higgs mass from quantum loops would be at the $n$-loop level of order $\Delta m_H^{(n)} \sim M_{\rm GUT}(\alpha/4\pi)^{n/2}$. %where $n$ is the number of loops.
For $m_H \sim 125$ GeV, the tuning must be done after $n=12$ loop corrections are included.  We note that this is nevertheless only one fine-tuning, albeit not easily enforceable by actual calculations.  The Hermitian Higgs doublet mass matrix is tuned to have near-zero determinant.  In contrast, in minimal SUSY GUTs, the needed tree level tuning requires the determinant of the complex doublet Higgsino mass matrix to be near zero.  Recall that in SUSY all mass parameters in the superpotential are complex in general.  Such a tuning amounts to two conditions, unlike the non-supersymmetric tuning, which requires only one such condition.  Although the Higgs mass can be ensured only after including very high order corrections in non-SUSY $SO(10)$, we find it intriguing that the fine-tuning condition is more minimal here compared to minimal SUSY $SO(10)$.

\section{Proof of \boldmath{$m_\tau \simeq 3 m_b$} in models with \boldmath{$126_H$} and \boldmath{$2 \times 120_H$}}

Without loss of generality, we can diagonalize the Yukawa coupling of $126_H$.  We focus on the second and third generation down quarks and charged leptons.  Their mass matrices can be written down as
\begin{eqnarray}
M_D = \left(\begin{matrix}a & c \\ -c & b  \end{matrix} \right),~~~~M_E = \left(\begin{matrix}-3a & c' \\ -c' & -3b  \end{matrix} \right).
\end{eqnarray}
This form persists even when many of $120_H$ fields are used, with their mass contribution going into the off-diagonal entries differently in $M_D$ and $M_E$. The exact invariants of these matrices are then
\begin{eqnarray}
m_s^2 + m_b^2 &=& |a|^2 + |b|^2 + 2 \,|c|^2 \label{app} \\
m_\mu^2 + m_\tau^2 &=& 9\,(|a|^2 + |b|^2) + 2\, |c'|^2 \\
m_s m_b &=& |ab + c^2| \\
m_\mu m_\tau &=& |9\,ab + c'^2|
\end{eqnarray}
From these relations it follows that
\begin{equation}
m_\mu^2 + m_\tau^2 = 9 \,(m_s^2 + m_b^2) + 18\, \left[\left|c^2 - m_s m_b e^{i \alpha} + \frac{m_\mu m_\tau}{9} e^{i \beta}\right|  -|c|^2\right]~.
\end{equation}
The undetermined parameter $c$ is bounded by Eq. (\ref{app}), and no matter how we vary $c$, the deviation from 3 in the ratio $m_\tau/m_b$ is of order $m_s/m_b \sim 5\%$.  The inclusion of the first family is not expected to change considerably
this result. This proves that a Higgs sector consisting of one $126_H$ and two or any number of copies of real $120_H$ cannot lead to realistic fermion masses.

%%%%%%%%%%%%%%%%%%%%%%%%%%%%%%%%%%%%%%%
%%%%%%%%%%%%%%%%%%%%%%%%%%%%%%%%%%%%%%%
\section{Best fit parameters for type-I dominance scenario} \label{A}
In this appendix we present the parameter set corresponding to the $\chi^2$ best fit for the type-I scenario.  \footnote{To reproduce the observables  presented in Table. \ref{tab:01} for both the type-I and type-I+II scenarios, one must keep all the significant digits of the parameters presented in these appendices. This high level of accuracy is needed to reproduce the neutrino observables; it is due to the fact that the right-handed neutrino mass spectrum in both cases shows extreme hierarchy among the generations, see Table. \ref{tab:02}.  Since this hierarchy between the first and the second generations is extreme, chopping-off digits effects mainly the quantity $\Delta m^2_{sol}$. }.

\begin{align}
&r_1 = -3.5178190\times 10^{-3}-5.1827520\times 10^{-3} i  \\
&r_2= -1.0441669+1.6253165\times 10^{-1} i  \\
&\phi=-7.9459769\times 10^{-1}  \\
&c_R= 5.8035176\times 10^{12} \;\rm{GeV}
\end{align}

\begin{equation}
S= \left(
\begin{array}{ccc}
 1.7926501\times 10^{-8} & 0 & 0 \\
 0 & 2.1219581\times 10^{-1} & 0 \\
 0 & 0 & 7.4949627\times 10^1
\end{array}
\right) \;\rm{GeV}
\end{equation}

\begin{equation}
D= \left(
\scalemath{0.7}{
\begin{array}{ccc}
 2.8344746\times 10^{-4}-5.3097883\times 10^{-4} i &
   4.501669\times 10^{-3}-1.7083332\times 10^{-3} i &
   6.0343793\times 10^{-2}+7.8900202\times 10^{-3} i \\
 4.501669\times 10^{-3}-1.7083332\times 10^{-3} i &
   2.7783311\times 10^{-2}-1.5722435\times 10^{-2} i &
   3.0561540\times 10^{-1}+9.6327579\times 10^{-2} i \\
 6.0343793\times 10^{-2}+7.8900202\times 10^{-3} i &
   3.0561540\times 10^{-1}+9.6327579\times 10^{-2} i &
   -4.304058\times 10^{-1}+2.4126529\times 10^{-1}  i
\end{array}
}
\right) \;\rm{GeV}
\end{equation}

\begin{equation}
A=
\scalemath{0.7}{
\left(
\begin{array}{ccc}
 0 & -3.9710310\times 10^{-3}-1.6550999\times 10^{-3}  i &
   -4.0391236\times 10^{-2}-4.2504129\times 10^{-2} i \\
 3.9710310\times 10^{-3}+1.6550999\times 10^{-3} i & 0 &
   -1.7267986\times 10^{-1}-3.2019088\times 10^{-1} i \\
 4.0391236\times 10^{-2}+4.2504129\times 10^{-2} i &
   1.7267986\times 10^{-1}+3.2019088\times 10^{-1} i & 0
\end{array}
\right)
} \;\rm{GeV}
\end{equation}

%%%%%%%%%%%%%%%%%%%%%%%%%%%%%%%%%%%%%%%
%%%%%%%%%%%%%%%%%%%%%%%%%%%%%%%%%%%%%%%
\section{Best fit parameters for type-I+II scenario} \label{B}
In this appendix we present the parameter set corresponding to the $\chi^2$ best fit for the type-I+II scenario.

\begin{align}
&r_1 = 4.1628007\times 10^{-3}-3.1705843\times
   10^{-3} i\\
&r_2= -7.4367427\times 10^{-1}+3.5915531\times
   10^{-1} i \\
&\phi= -6.4632781\times 10^{-1} \\
&c_R= 4.2254013\times 10^{10}\;\rm{GeV} \\
&c_L= 1.5155879\times 10^{-10}-1.4499546\times
   10^{-11} i \;\rm{GeV}
\end{align}
\begin{equation}
S= \left(
\begin{array}{ccc}
 1.4547716\times 10^{-4} & 0 & 0 \\
 0 & 2.6693088\times 10^{-1} & 0 \\
 0 & 0 & 7.4473135\times 10^1
\end{array}
\right) \;\rm{GeV} \\
\end{equation}

\begin{equation}
D=
\scalemath{0.7}{
\left(
\begin{array}{ccc}
 4.8953934\times 10^{-4}-2.6113522\times
   10^{-4} i & -1.6504521\times
   10^{-5}+1.1420336\times 10^{-2} i &
   -2.151214\times 10^{-1}+1.7234983\times
   10^{-2} i \\
 -1.6504521\times 10^{-5}+1.1420336\times
   10^{-2} i & -2.8562186\times
   10^{-2}+2.8403787\times 10^{-2} i &
   -3.7065300\times 10^{-1}-2.0521574\times
   10^{-1} i \\
 -2.151214\times 10^{-1}+1.7234983\times
   10^{-2} i & -3.7065300\times
   10^{-1}-2.0521574\times 10^{-1} i &
   3.6722700\times 10^{-2}+2.6598904\times
   10^{-1} i
\end{array}
\right)
}
 \;\rm{GeV}
\end{equation} \\
\begin{equation}
A=
\scalemath{0.7}{
\left(
\begin{array}{ccc}
 0 & 2.518929\times 10^{-3}-1.1393329\times
   10^{-2} i & 1.7915567\times
   10^{-1}+1.1538080\times 10^{-1} i \\
 -2.518929\times 10^{-3}+1.1393329\times
   10^{-2} i & 0 & 1.6923025\times
   10^{-1}+3.6425489\times 10^{-1} i \\
 -1.7915567\times 10^{-1}-1.1538080\times
   10^{-1} i & -1.6923025\times
   10^{-1}-3.6425489\times 10^{-1} i & 0
\end{array}
\right)
} \;\rm{GeV}
\end{equation}

\end{appendices}

%%%%%%%%%%%%%%%%%%%%%%%%%%%%%%%%%%%%%%%%%%%%%
%%%%%%%%%%%%%%%%%%%%%%%%%%%%%%%%%%%%%%%%%%%%%
%\newpage
\FloatBarrier


\begin{thebibliography}{99}

%\cite{Pati:1974yy}
\bibitem{Pati:1974yy}
  J.~C.~Pati and A.~Salam,
  ``Lepton Number as the Fourth Color,''
  Phys.\ Rev.\ D {\bf 10}, 275 (1974)
  Erratum: [Phys.\ Rev.\ D {\bf 11}, 703 (1975)].
 % doi:10.1103/PhysRevD.10.275, 10.1103/PhysRevD.11.703.2
  %%CITATION = doi:10.1103/PhysRevD.10.275, 10.1103/PhysRevD.11.703.2;%%
  %4099 citations counted in INSPIRE as of 12 Nov 2016%

%  \cite{Georgi:1974sy}
\bibitem{Georgi:1974sy}
  H.~Georgi and S.~L.~Glashow,
  ``Unity of All Elementary Particle Forces,''
  Phys.\ Rev.\ Lett.\  {\bf 32}, 438 (1974).
 % doi:10.1103/PhysRevLett.32.438
  %%CITATION = doi:10.1103/PhysRevLett.32.438;%%
  %4262 citations counted in INSPIRE as of 12 Nov 2016

%\cite{Georgi:1974yf}
\bibitem{Georgi:1974yf}
  H.~Georgi, H.~R.~Quinn and S.~Weinberg,
  ``Hierarchy of Interactions in Unified Gauge Theories,''
  Phys.\ Rev.\ Lett.\  {\bf 33}, 451 (1974).
 % doi:10.1103/PhysRevLett.33.451
  %%CITATION = doi:10.1103/PhysRevLett.33.451;%%
  %1741 citations counted in INSPIRE as of 12 Nov 2016

\bibitem{so10}
H. Georgi, in Particles and Fields (edited by C.E. Carlson), A.I.P. (1975);
%\cite{Fritzsch:1974nn}
%\bibitem{Fritzsch:1974nn}
  H.~Fritzsch and P.~Minkowski,
  ``Unified Interactions of Leptons and Hadrons,''
  Annals Phys.\  {\bf 93} (1975) 193.
%  doi:10.1016/0003-4916(75)90211-0
  %%CITATION = doi:10.1016/0003-4916(75)90211-0;%%
  %1528 citations counted in INSPIRE as of 13 Dec 2016



%%%%%%%%%%%%%%%%%%%%%%%%
%%%%% TYPE I %%%%%%%%%%%%%%
\bibitem{Minkowski:1977sc}
P.~Minkowski,
Phys.\ Lett.\ B {\bf 67} (1977) 421;
%%CITATION = PHLTA,B67,421;%%
T.~Yanagida, proceedings of the {\em Workshop on Unified Theories
and Baryon Number in the Universe}, Tsukuba, 1979, eds.
A. Sawada, A. Sugamoto;
S.~Glashow, in {\em Cargese 1979, Proceedings, Quarks and Leptons}
(1979);
M.~Gell-Mann, P.~Ramond, R.~Slansky, proceedings of the
{\em Supergravity Stony Brook Workshop}, New York, 1979,
eds. P. Van Niewenhuizen, D. Freeman;
R.~Mohapatra, G.~Senjanovi\' c,
Phys.Rev.Lett. {\bf 44} (1980) 912.
%%CITATION = PRLTA,44,912;%%

%%%%%%%%%%%%%%%%%%%%%%%%
%%%%% TYPE II %%%%%%%%%%%%%%
%\cite{Schechter:1980gr}
\bibitem{Schechter:1980gr}
  J.~Schechter and J.~W.~F.~Valle,
  ``Neutrino Masses in SU(2) x U(1) Theories,''
  Phys.\ Rev.\ D {\bf 22}, 2227 (1980);
 % doi:10.1103/PhysRevD.22.2227
  %%CITATION = doi:10.1103/PhysRevD.22.2227;%%
  %1982 citations counted in INSPIRE as of 06 Dec 2016
  %\cite{Schechter:1981cv}
%\cite{Lazarides:1980nt}
%\bibitem{Lazarides:1980nt}
  G.~Lazarides, Q.~Shafi and C.~Wetterich,
  ``Proton Lifetime and Fermion Masses in an SO(10) Model,''
  Nucl.\ Phys.\ B {\bf 181}, 287 (1981);
  %doi:10.1016/0550-3213(81)90354-0
  %%CITATION = doi:10.1016/0550-3213(81)90354-0;%%
  %1043 citations counted in INSPIRE as of 12 Nov 2016
  R.~N.~Mohapatra and G.~Senjanovi\' c,
  ``Neutrino Masses and Mixings in Gauge Models with Spontaneous Parity Violation,''
  Phys.\ Rev.\ D {\bf 23}, 165 (1981);
  %doi:10.1103/PhysRevD.23.165
  %%CITATION = doi:10.1103/PhysRevD.23.165;%%
  %2034 citations counted in INSPIRE as of 06 Dec 2016
  %\cite{Wetterich:1981bx}
%\bibitem{Wetterich:1981bx}
  C.~Wetterich,
  ``Neutrino Masses and the Scale of B-L Violation,''
  Nucl.\ Phys.\ B {\bf 187}, 343 (1981);
  %doi:10.1016/0550-3213(81)90279-0
  %%CITATION = doi:10.1016/0550-3213(81)90279-0;%%
  %307 citations counted in INSPIRE as of 06 Dec 2016
%\bibitem{Schechter:1981cv}
  J.~Schechter and J.~W.~F.~Valle,
  ``Neutrino Decay and Spontaneous Violation of Lepton Number,''
  Phys.\ Rev.\ D {\bf 25}, 774 (1982).
  %doi:10.1103/PhysRevD.25.774
  %%CITATION = doi:10.1103/PhysRevD.25.774;%%
  %651 citations counted in INSPIRE as of 06 Dec 2016
 %\cite{Mohapatra:1980yp}
%\bibitem{Mohapatra:1980yp}










\bibitem{GoranIntSc:1981}
%\cite{Rizzo:1981su}
%\bibitem{Rizzo:1981su}
  T.~G.~Rizzo and G.~Senjanovi\' c,
  ``Can There Be Low Intermediate Mass Scales in Grand Unified Theories?,''
  Phys.\ Rev.\ Lett.\  {\bf 46}, 1315 (1981);
  %%CITATION = PRLTA,46,1315;%%
  %135 citations counted in INSPIRE as of 15 Aug 2015
  %\cite{Rizzo:1981dm}
%\bibitem{Rizzo:1981dm}
  T.~G.~Rizzo and G.~Senjanovi\' c,
  ``Grand Unification and Parity Restoration at Low-Energies. 1. Phenomenology,''
  Phys.\ Rev.\ D {\bf 24}, 704 (1981)
  [Phys.\ Rev.\ D {\bf 25}, 1447 (1982)];
  %%CITATION = PHRVA,D24,704;%%
  %169 citations counted in INSPIRE as of 15 Aug 2015
  %\cite{Rizzo:1981jr}
%\bibitem{Rizzo:1981jr}
  T.~G.~Rizzo and G.~Senjanovi\' c,
  ``Grand Unification and Parity Restoration at Low-energies. 2. Unification Constraints,''
  Phys.\ Rev.\ D {\bf 25}, 235 (1982).
  %%CITATION = PHRVA,D25,235;%%
  %85 citations counted in INSPIRE as of 15 Aug 2015

%\cite{Caswell:1982fx}
\bibitem{Caswell:1982fx}
 W.~E.~Caswell, J.~Milutinovi\' c, and G.~Senjanovi\' c,
  ``Predictions of Left-right Symmetric Grand Unified Theories,''
  Phys.\ Rev.\ D {\bf 26}, 161 (1982).
  %%CITATION = PHRVA,D26,161;%%
  %18 citations counted in INSPIRE as of 15 Aug 2015


%\cite{Chang:1983fu}
\bibitem{Chang:1983fu}
  D.~Chang, R.~N.~Mohapatra and M.~K.~Parida,
  ``Decoupling Parity and SU(2)-R Breaking Scales: A New Approach to Left-Right Symmetric Models,''
  Phys.\ Rev.\ Lett.\  {\bf 52}, 1072 (1984).
 % doi:10.1103/PhysRevLett.52.1072
  %%CITATION = doi:10.1103/PhysRevLett.52.1072;%%
  %241 citations counted in INSPIRE as of 13 Nov 2016

%\cite{Gipson:1984aj}
\bibitem{Gipson:1984aj}
  J.~M.~Gipson and R.~E.~Marshak,
  ``Intermediate Mass Scales in the New SO(10) Grand Unification in the One Loop Approximation,''
  Phys.\ Rev.\ D {\bf 31}, 1705 (1985).
  %%CITATION = PHRVA,D31,1705;%%
  %41 citations counted in INSPIRE as of 23 juin 2015

  %\cite{Chang:1984qr}
\bibitem{Chang:1984qr}
  D.~Chang, R.~N.~Mohapatra, J.~Gipson, R.~E.~Marshak and M.~K.~Parida,
  ``Experimental Tests of New SO(10) Grand Unification,''
  Phys.\ Rev.\ D {\bf 31}, 1718 (1985).
  %%CITATION = PHRVA,D31,1718;%%
  %162 citations counted in INSPIRE as of 23 juin 2015



  %\cite{Deshpande:1992au}
\bibitem{Deshpande:1992au}
  N.~G.~Deshpande, E.~Keith and P.~B.~Pal,
  ``Implications of LEP results for SO(10) grand unification,''
  Phys.\ Rev.\ D {\bf 46}, 2261 (1993).
  %%CITATION = PHRVA,D46,2261;%%
  %86 citations counted in INSPIRE as of 23 Jun 2015

  %\cite{Deshpande:1992em}
\bibitem{Deshpande:1992em}
  N.~G.~Deshpande, E.~Keith and P.~B.~Pal,
  ``Implications of LEP results for SO(10) grand unification with two intermediate stages,''
  Phys.\ Rev.\ D {\bf 47}, 2892 (1993)
  [hep-ph/9211232].
  %%CITATION = HEP-PH/9211232;%%
  %37 citations counted in INSPIRE as of 23 Jun 2015

%\cite{Bertolini:2009qj}
\bibitem{Bertolini:2009qj}
  S.~Bertolini, L.~Di Luzio and M.~Malinsk\'y,
  ``Intermediate mass scales in the non-supersymmetric SO(10) grand unification: A Reappraisal,''
  Phys.\ Rev.\ D {\bf 80}, 015013 (2009)
  [arXiv:0903.4049 [hep-ph]].
  %%CITATION = ARXIV:0903.4049;%%
  %65 citations counted in INSPIRE as of 23 juin 2015

  %%%%%%%%%%%%%%%%%%%%%%%%%%%%%%%%%%%
%%%%%%%%%%%   45+126   %%%%%%%%%%%%
%\cite{Bertolini:2009es}
\bibitem{Bertolini:2009es}
  S.~Bertolini, L.~Di Luzio and M.~Malinsk\'y,
  ``On the vacuum of the minimal nonsupersymmetric SO(10) unification,''
  Phys.\ Rev.\ D {\bf 81}, 035015 (2010)
  % doi:10.1103/PhysRevD.81.035015,
  [arXiv:0912.1796 [hep-ph]].
  %%CITATION = doi:10.1103/PhysRevD.81.035015;%%
  %37 citations counted in INSPIRE as of 15 May 2016

%\cite{Bertolini:2010ng}
\bibitem{Bertolini:2010ng}
  S.~Bertolini, L.~Di Luzio and M.~Malinsk\'y,
  ``The quantum vacuum of the minimal SO(10) GUT,''
  J.\ Phys.\ Conf.\ Ser.\  {\bf 259}, 012098 (2010)
  % doi:10.1088/1742-6596/259/1/012098,
  [arXiv:1010.0338 [hep-ph]].
  %%CITATION = doi:10.1088/1742-6596/259/1/012098;%%
  %5 citations counted in INSPIRE as of 15 May 2016

  %%%%%%%%%%%%%%%%%%%%%%%%%%%%%%%%%%%%%%%%%
%%%%%%%%%%%%%%%   54+126  %%%%%%%%%%%%%%%%%%%%%
\bibitem{khan}
 K.~S.~Babu and S.~Khan,
  ``Minimal nonsupersymmetric $SO(10)$ model: Gauge coupling unification, proton decay, and fermion masses,''
  Phys.\ Rev.\ D {\bf 92}, no. 7, 075018 (2015)
 %  doi:10.1103/PhysRevD.92.075018
  [arXiv:1507.06712 [hep-ph]].
  %%CITATION = doi:10.1103/PhysRevD.92.075018;%%
  %6 citations counted in INSPIRE as of 16 May 2016

%\cite{Graf:2016znk}
\bibitem{Graf:2016znk}
L.~Gr\'af, M.~Malinsk\'y, T.~Mede and V.~Susi\v c,
``One-Loop Pseudo-Goldstone Masses in the Minimal $SO(10)$ Higgs Model,''
arXiv:1611.01021 [hep-ph].
%%CITATION = ARXIV:1611.01021;%%

\bibitem{susyunification}
%\cite{Dimopoulos:1981yj}
%\bibitem{Dimopoulos:1981yj}
  S.~Dimopoulos, S.~Raby and F.~Wilczek,
  ``Supersymmetry and the Scale of Unification,''
  Phys.\ Rev.\ D {\bf 24}, 1681 (1981);
 % doi:10.1103/PhysRevD.24.1681
  %%CITATION = doi:10.1103/PhysRevD.24.1681;%%
  %1025 citations counted in INSPIRE as of 21 Sep 2016
%\cite{Ibanez:1981yh}
%\bibitem{Ibanez:1981yh}
  L.~E.~Ib\'a\~nez and G.~G.~Ross,
  ``Low-Energy Predictions in Supersymmetric Grand Unified Theories,''
  Phys.\ Lett.\ B {\bf 105}, 439 (1981);
  %doi:10.1016/0370-2693(81)91200-4
  %%CITATION = doi:10.1016/0370-2693(81)91200-4;%%
  %543 citations counted in INSPIRE as of 21 Sep 2016
%\cite{Einhorn:1981sx}
%\bibitem{Einhorn:1981sx}
  M.~B.~Einhorn and D.~R.~T.~Jones,
  ``The Weak Mixing Angle and Unification Mass in Supersymmetric SU(5),''
  Nucl.\ Phys.\ B {\bf 196}, 475 (1982);
 % doi:10.1016/0550-3213(82)90502-8
  %%CITATION = doi:10.1016/0550-3213(82)90502-8;%%
  %585 citations counted in INSPIRE as of 21 Sep 2016
%\cite{Marciano:1981un}
%\bibitem{Marciano:1981un}
  W.~J.~Marciano and G.~Senjanovi\' c,
  ``Predictions of Supersymmetric Grand Unified Theories,''
  Phys.\ Rev.\ D {\bf 25}, 3092 (1982).
  %doi:10.1103/PhysRevD.25.3092
  %%CITATION = doi:10.1103/PhysRevD.25.3092;%%
  %345 citations counted in INSPIRE as of 21 Sep 2016

  \bibitem{bb}
  K.~S.~Babu and S.~M.~Barr,
  ``An SO(10) solution to the puzzle of quark and lepton masses,''
  Phys.\ Rev.\ Lett.\  {\bf 75}, 2088 (1995)
  %doi:10.1103/PhysRevLett.75.2088
  [hep-ph/9503215].
  %%CITATION = doi:10.1103/PhysRevLett.75.2088;%%
  %43 citations counted in INSPIRE as of 07 Dec 2016

  \bibitem{bbs}
   K.~S.~Babu, B.~Bajc and S.~Saad,
  ``New Class of SO(10) Models for Flavor,''
  Phys.\ Rev.\ D {\bf 94}, no. 1, 015030 (2016)
 % doi:10.1103/PhysRevD.94.015030
  [arXiv:1605.05116 [hep-ph]].

  %\cite{Lazarides:1980nt}
\bibitem{Lazarides:1980nt}
  G.~Lazarides, Q.~Shafi and C.~Wetterich,
  ``Proton Lifetime and Fermion Masses in an SO(10) Model,''
  Nucl.\ Phys.\ B {\bf 181}, 287 (1981).
%  doi:10.1016/0550-3213(81)90354-0
  %%CITATION = doi:10.1016/0550-3213(81)90354-0;%%
  %1049 citations counted in INSPIRE as of 11 Dec 2016

  %\cite{Davidson:1983fe}
\bibitem{Davidson:1983fe}
  A.~Davidson, V.~P.~Nair and K.~C.~Wali,
  ``Mixing Angles and {CP} Violation in the $SO(10) \times U(1)_{(pq)}$ Model,''
  Phys.\ Rev.\ D {\bf 29}, 1513 (1984).
 % doi:10.1103/PhysRevD.29.1513
  %%CITATION = doi:10.1103/PhysRevD.29.1513;%%
  %35 citations counted in INSPIRE as of 11 Dec 2016

  %\cite{Lazarides:1990ni}
\bibitem{Lazarides:1990ni}
  G.~Lazarides and Q.~Shafi,
  ``Fermion masses and mixings in SO(10),''
  Nucl.\ Phys.\ B {\bf 350}, 179 (1991).
 % doi:10.1016/0550-3213(91)90257-X
  %%CITATION = doi:10.1016/0550-3213(91)90257-X;%%
  %44 citations counted in INSPIRE as of 11 Dec 2016

  %\cite{Babu:1992ia}
\bibitem{Babu:1992ia}
  K.~S.~Babu and R.~N.~Mohapatra,
  ``Predictive neutrino spectrum in minimal SO(10) grand unification,''
  Phys.\ Rev.\ Lett.\  {\bf 70}, 2845 (1993)
  %doi:10.1103/PhysRevLett.70.2845
  [hep-ph/9209215].
  %%CITATION = doi:10.1103/PhysRevLett.70.2845;%%
  %359 citations counted in INSPIRE as of 12 Nov 2016

  %\cite{Lee:1994qx}
\bibitem{Lee:1994qx}
  D.~G.~Lee and R.~N.~Mohapatra,
  ``An SO(10) x S(4) scenario for naturally degenerate neutrinos,''
  Phys.\ Lett.\ B {\bf 329}, 463 (1994)
%  doi:10.1016/0370-2693(94)91091-X
  [hep-ph/9403201].
  %%CITATION = doi:10.1016/0370-2693(94)91091-X;%%
  %124 citations counted in INSPIRE as of 11 Dec 2016


%\cite{Bajc:2005zf}
\bibitem{Bajc:2005zf}
  B.~Bajc, A.~Melfo, G.~Senjanovi\' c and F.~Vissani,
  ``Yukawa sector in non-supersymmetric renormalizable SO(10),''
  Phys.\ Rev.\ D {\bf 73}, 055001 (2006)
  %doi:10.1103/PhysRevD.73.055001
  [hep-ph/0510139].
  %%CITATION = doi:10.1103/PhysRevD.73.055001;%%
  %62 citations counted in INSPIRE as of 13 Nov 2016

%\cite{Joshipura:2011nn}
\bibitem{Joshipura:2011nn}
  A.~S.~Joshipura and K.~M.~Patel,
  ``Fermion Masses in SO(10) Models,''
  Phys.\ Rev.\ D {\bf 83}, 095002 (2011)
 % doi:10.1103/PhysRevD.83.095002
  [arXiv:1102.5148 [hep-ph]].
  %%CITATION = doi:10.1103/PhysRevD.83.095002;%%
  %54 citations counted in INSPIRE as of 28 Nov 2016


%\cite{Altarelli:2013aqa}
\bibitem{Altarelli:2013aqa}
  G.~Altarelli and D.~Meloni,
  ``A non supersymmetric SO(10) grand unified model for all the physics below $M_{GUT}$,''
  JHEP {\bf 1308}, 021 (2013)
 % doi:10.1007/JHEP08(2013)021
  [arXiv:1305.1001 [hep-ph]].
  %%CITATION = doi:10.1007/JHEP08(2013)021;%%
  %30 citations counted in INSPIRE as of 06 Dec 2016


%\cite{Dueck:2013gca}
\bibitem{Dueck:2013gca}
  A.~Dueck and W.~Rodejohann,
  ``Fits to SO(10) Grand Unified Models,''
  JHEP {\bf 1309}, 024 (2013)
  [arXiv:1306.4468 [hep-ph]].
  %%CITATION = ARXIV:1306.4468;%%
  %12 citations counted in INSPIRE as of 30 Jun 2015








%%%%%%%%%%%%%%%%%%%%%%%%%%%%%%%%%%%%%%%
%%%%%%%%%%%  10+126   %%%%%%%%%%%%%%%

%\cite{Bajc:2001fe}
\bibitem{Bajc:2001fe}
  B.~Bajc, G.~Senjanovi\' c and F.~Vissani,
  ``How neutrino and charged fermion masses are connected within minimal supersymmetric SO(10),''
  PoS HEP {\bf 2001}, 198 (2001)
  [hep-ph/0110310].
  %%CITATION = HEP-PH/0110310;%%
  %68 citations counted in INSPIRE as of 06 Dec 2016



%\cite{Fukuyama:2002ch}
\bibitem{Fukuyama:2002ch}
  T.~Fukuyama and N.~Okada,
  ``Neutrino oscillation data versus minimal supersymmetric SO(10) model,''
  JHEP {\bf 0211}, 011 (2002)
  [hep-ph/0205066].
  %%CITATION = HEP-PH/0205066;%%
  %139 citations counted in INSPIRE as of 30 Jun 2015

%\cite{Bajc:2002iw}
\bibitem{Bajc:2002iw}
  B.~Bajc, G.~Senjanovi\' c and F.~Vissani,
  ``b - tau unification and large atmospheric mixing: A Case for noncanonical seesaw,''
  Phys.\ Rev.\ Lett.\  {\bf 90}, 051802 (2003)
  [hep-ph/0210207].
  %%CITATION = HEP-PH/0210207;%%
  %236 citations counted in INSPIRE as of 30 juin 2015

%\cite{Goh:2003sy}
\bibitem{Goh:2003sy}
  H.~S.~Goh, R.~N.~Mohapatra and S.~P.~Ng,
  ``Minimal SUSY SO(10), b tau unification and large neutrino mixings,''
  Phys.\ Lett.\ B {\bf 570}, 215 (2003)
  [hep-ph/0303055].
  %%CITATION = HEP-PH/0303055;%%
  %208 citations counted in INSPIRE as of 30 Jun 2015

%\cite{Goh:2003hf}
\bibitem{Goh:2003hf}
  H.~S.~Goh, R.~N.~Mohapatra and S.~P.~Ng,
  ``Minimal SUSY SO(10) model and predictions for neutrino mixings and leptonic CP violation,''
  Phys.\ Rev.\ D {\bf 68}, 115008 (2003)
  [hep-ph/0308197].
  %%CITATION = HEP-PH/0308197;%%
  %184 citations counted in INSPIRE as of 30 juin 2015


%\cite{Babu:2005ia}
\bibitem{Babu:2005ia}
  K.~S.~Babu and C.~Macesanu,
  ``Neutrino masses and mixings in a minimal SO(10) model,''
  Phys.\ Rev.\ D {\bf 72}, 115003 (2005)
  [hep-ph/0505200].
  %%CITATION = HEP-PH/0505200;%%
  %105 citations counted in INSPIRE as of 11 juin 2015

  %\cite{Bertolini:2006pe}
\bibitem{Bertolini:2006pe}
  S.~Bertolini, T.~Schwetz and M.~Malinsk\'y,
  ``Fermion masses and mixings in SO(10) models and the neutrino challenge to SUSY GUTs,''
  Phys.\ Rev.\ D {\bf 73}, 115012 (2006)
%  doi:10.1103/PhysRevD.73.115012
  [hep-ph/0605006].
  %%CITATION = doi:10.1103/PhysRevD.73.115012;%%
  %108 citations counted in INSPIRE as of 11 Dec 2016




%\cite{Bajc:2008dc}
\bibitem{Bajc:2008dc}
B.~Bajc, I.~Dor\v sner and M.~Nemev\v sek,
``Minimal SO(10) Splits Supersymmetry,''
JHEP {\bf 0811} (2008) 007
%doi:10.1088/1126-6708/2008/11/007
[arXiv:0809.1069 [hep-ph]].
%%CITATION = doi:10.1088/1126-6708/2008/11/007;%%<br /> 44 citations counted in INSPIRE as of 05 Dec 2016

%\cite{Fukuyama:2015kra}
\bibitem{Fukuyama:2015kra}
  T.~Fukuyama, K.~Ichikawa and Y.~Mimura,
  ``Revisiting fermion mass and mixing fits in the minimal SUSY $SO(10)$ GUT,''
  Phys.\ Rev.\ D {\bf 94}, no. 7, 075018 (2016)
%  doi:10.1103/PhysRevD.94.075018
  [arXiv:1508.07078 [hep-ph]].
  %%CITATION = doi:10.1103/PhysRevD.94.075018;%%
  %8 citations counted in INSPIRE as of 13 Dec 2016

  %\cite{Bertolini:2004eq}
\bibitem{Bertolini:2004eq}
  S.~Bertolini, M.~Frigerio and M.~Malinsk\'y,
  ``Fermion masses in SUSY SO(10) with type II seesaw: A Non-minimal predictive scenario,''
  Phys.\ Rev.\ D {\bf 70}, 095002 (2004)
  [hep-ph/0406117].
  %%CITATION = HEP-PH/0406117;%%

 %\cite{Yang:2004xt}
\bibitem{Yang:2004xt}
  W.~M.~Yang and Z.~G.~Wang,
  ``Fermion masses and flavor mixing in a supersymmetric SO(10) model,''
  Nucl.\ Phys.\ B {\bf 707}, 87 (2005)
%  doi:10.1016/j.nuclphysb.2004.11.042
  [hep-ph/0406221].
  %%CITATION = doi:10.1016/j.nuclphysb.2004.11.042;%%
  %45 citations counted in INSPIRE as of 13 Dec 2016


    %\cite{Dutta:2004zh}
\bibitem{Dutta:2004zh}
  B.~Dutta, Y.~Mimura and R.~N.~Mohapatra,
  ``Suppressing proton decay in the minimal SO(10) model,''
  Phys.\ Rev.\ Lett.\  {\bf 94}, 091804 (2005)
%  doi:10.1103/PhysRevLett.94.091804
  [hep-ph/0412105].
  %%CITATION = doi:10.1103/PhysRevLett.94.091804;%%
  %79 citations counted in INSPIRE as of 11 Dec 2016

%\cite{Aulakh:2008sn}
\bibitem{Aulakh:2008sn}
  C.~S.~Aulakh and S.~K.~Garg,
  ``The New Minimal Supersymmetric GUT : Spectra, RG analysis and Fermion Fits,''
  Nucl.\ Phys.\ B {\bf 857}, 101 (2012)
%  doi:10.1016/j.nuclphysb.2011.12.003
  [arXiv:0807.0917 [hep-ph]].
  %%CITATION = doi:10.1016/j.nuclphysb.2011.12.003;%%
  %46 citations counted in INSPIRE as of 11 Dec 2016

    %\cite{Chen:2003zv}
\bibitem{Chen:2003zv}
  M.~C.~Chen and K.~T.~Mahanthappa,
  ``Fermion masses and mixing and CP violation in SO(10) models with family symmetries,''
  Int.\ J.\ Mod.\ Phys.\ A {\bf 18}, 5819 (2003)
%  doi:10.1142/S0217751X03017026
  [hep-ph/0305088].
  %%CITATION = doi:10.1142/S0217751X03017026;%%
  %102 citations counted in INSPIRE as of 13 Dec 2016

  %\cite{Grimus:2006bb}
\bibitem{Grimus:2006bb}
  W.~Grimus and H.~Kuhbock,
  ``Fermion masses and mixings in a renormalizable $SO(10) \times Z(2)$ GUT,''
  Phys.\ Lett.\ B {\bf 643}, 182 (2006)
 % doi:10.1016/j.physletb.2006.10.038
  [hep-ph/0607197].
  %%CITATION = doi:10.1016/j.physletb.2006.10.038;%%
  %52 citations counted in INSPIRE as of 13 Dec 2016

  %\cite{Cai:2006mf}
\bibitem{Cai:2006mf}
  Y.~Cai and H.~B.~Yu,
  ``A SO(10) GUT Model with S4 Flavor Symmetry,''
  Phys.\ Rev.\ D {\bf 74}, 115005 (2006)
%  doi:10.1103/PhysRevD.74.115005
  [hep-ph/0608022].
  %%CITATION = doi:10.1103/PhysRevD.74.115005;%%
  %81 citations counted in INSPIRE as of 13 Dec 2016


  %\cite{Albaid:2011vr}
\bibitem{Albaid:2011vr}
  A.~Albaid,
  ``Flavor Violation in a Minimal $SO(10)\times A_4$ SUSY GUT,''
  Int.\ J.\ Mod.\ Phys.\ A {\bf 27}, 1250005 (2012)
 % doi:10.1142/S0217751X12500054
  [arXiv:1106.4070 [hep-ph]].
  %%CITATION = doi:10.1142/S0217751X12500054;%%
  %2 citations counted in INSPIRE as of 13 Dec 2016

    %\cite{Ferreira:2015jpa}
\bibitem{Ferreira:2015jpa}
  P.~M.~Ferreira, W.~Grimus, D.~Jur\v ciukonis and L.~Lavoura,
  ``Flavour symmetries in a renormalizable SO(10) model,''
  Nucl.\ Phys.\ B {\bf 906}, 289 (2016)
 % doi:10.1016/j.nuclphysb.2016.03.011
  [arXiv:1510.02641 [hep-ph]].
  %%CITATION = doi:10.1016/j.nuclphysb.2016.03.011;%%
  %3 citations counted in INSPIRE as of 13 Dec 2016

   %%%%%%%%%%%%%%%%%%
%%%%% neutrino mass via 16_H %%%%%%
%\cite{Witten:1979nr}
\bibitem{witten}
  E.~Witten,
  ``Neutrino Masses in the Minimal O(10) Theory,''
  Phys.\ Lett.\  {\bf 91B}, 81 (1980).
 % doi:10.1016/0370-2693(80)90666-8
  %%CITATION = doi:10.1016/0370-2693(80)90666-8;%%
  %367 citations counted in INSPIRE as of 06 Dec 2016


  \bibitem{bs}
     B.~Bajc and G.~Senjanovi\' c,
  ``Radiative seesaw: A Case for split supersymmetry,''
  Phys.\ Lett.\ B {\bf 610}, 80 (2005)
%  doi:10.1016/j.physletb.2005.01.074
  [hep-ph/0411193].

%\cite{Giudice:2011cg}
\bibitem{Giudice:2011cg}
  G.~F.~Giudice and A.~Strumia,
  ``Probing High-Scale and Split Supersymmetry with Higgs Mass Measurements,''
  Nucl.\ Phys.\ B {\bf 858}, 63 (2012)
 % doi:10.1016/j.nuclphysb.2012.01.001
  [arXiv:1108.6077 [hep-ph]].
  %%CITATION = doi:10.1016/j.nuclphysb.2012.01.001;%%
  %209 citations counted in INSPIRE as of 13 Dec 2016
%

%\cite{Antusch:2013jca}
\bibitem{Antusch:2013jca}
  S.~Antusch and V.~Maurer,
  ``Running quark and lepton parameters at various scales,''
  JHEP {\bf 1311}, 115 (2013)
  % doi:10.1007/JHEP11(2013)115,
  [arXiv:1306.6879 [hep-ph]].
  %%CITATION = doi:10.1007/JHEP11(2013)115;%%
  %31 citations counted in INSPIRE as of 15 May 2016

%\cite{Machacek:1983fi}
\bibitem{Machacek:1983fi}
  M.~E.~Machacek and M.~T.~Vaughn,
  ``Two Loop Renormalization Group Equations in a General Quantum Field Theory. 2. Yukawa Couplings,''
  Nucl.\ Phys.\ B {\bf 236}, 221 (1984);
 % doi:10.1016/0550-3213(84)90533-9
  %%CITATION = doi:10.1016/0550-3213(84)90533-9;%%
  %458 citations counted in INSPIRE as of 08 Jun 2016
%\cite{Arason:1991ic}
%\bibitem{Arason:1991ic}
  H.~Arason, D.~J.~Casta\~no, B.~Keszthelyi, S.~Mikaelian, E.~J.~Piard, P.~Ramond and B.~D.~Wright,
  ``Renormalization group study of the standard model and its extensions. 1. The Standard model,''
  Phys.\ Rev.\ D {\bf 46}, 3945 (1992).
  %doi:10.1103/PhysRevD.46.3945
  %%CITATION = doi:10.1103/PhysRevD.46.3945;%%
  %365 citations counted in INSPIRE as of 08 Jun 2016

%\cite{Babu:1987im}
\bibitem{Babu:1987im}
  K.~S.~Babu,
  ``Renormalization Group Analysis of the {Kobayashi-Maskawa} Matrix,''
  Z.\ Phys.\ C {\bf 35}, 69 (1987).
 % doi:10.1007/BF01561056
  %%CITATION = doi:10.1007/BF01561056;%%
  %82 citations counted in INSPIRE as of 08 Jun 2016


%\cite{Fogli:2012ua}
\bibitem{Fogli:2012ua}
  G.~L.~Fogli, E.~Lisi, A.~Marrone, D.~Montanino, A.~Palazzo and A.~M.~Rotunno,
  ``Global analysis of neutrino masses, mixings and phases: entering the era of leptonic CP violation searches,''
  Phys.\ Rev.\ D {\bf 86}, 013012 (2012)
  % doi:10.1103/PhysRevD.86.013012,
  [arXiv:1205.5254 [hep-ph]].
  %%CITATION = doi:10.1103/PhysRevD.86.013012;%%
  %611 citations counted in INSPIRE as of 15 May 2016

%\cite{Babu:1993qv}
\bibitem{Babu:1993qv}
  K.~S.~Babu, C.~N.~Leung and J.~T.~Pantaleone,
  ``Renormalization of the neutrino mass operator,''
  Phys.\ Lett.\ B {\bf 319}, 191 (1993)
  %doi:10.1016/0370-2693(93)90801-N
  [hep-ph/9309223];
  %%CITATION = doi:10.1016/0370-2693(93)90801-N;%%
  %335 citations counted in INSPIRE as of 06 Nov 2016
%\cite{Antusch:2001ck}
%\bibitem{Antusch:2001ck}
  S.~Antusch, M.~Drees, J.~Kersten, M.~Lindner and M.~Ratz,
  ``Neutrino mass operator renormalization revisited,''
  Phys.\ Lett.\ B {\bf 519}, 238 (2001)
  %doi:10.1016/S0370-2693(01)01127-3
  [hep-ph/0108005].
  %%CITATION = doi:10.1016/S0370-2693(01)01127-3;%%
  %194 citations counted in INSPIRE as of 06 Nov 2016

%\cite{Georgi:1979md}
\bibitem{Georgi:1979md}
  H.~Georgi,
  ``Towards a Grand Unified Theory of Flavor,''
  Nucl.\ Phys.\ B {\bf 156}, 126 (1979).
  %doi:10.1016/0550-3213(79)90497-8
  %%CITATION = doi:10.1016/0550-3213(79)90497-8;%%
  %445 citations counted in INSPIRE as of 16 Nov 2016
%\cite{delAguila:1980qag}
\bibitem{delAguila:1980qag}
  F.~del Aguila and L.~E.~Ib\'a\~nez,
  ``Higgs Bosons in SO(10) and Partial Unification,''
  Nucl.\ Phys.\ B {\bf 177}, 60 (1981).
  %doi:10.1016/0550-3213(81)90266-2
  %%CITATION = doi:10.1016/0550-3213(81)90266-2;%%
  %165 citations counted in INSPIRE as of 16 Nov 2016
%\cite{Mohapatra:1982aq}
\bibitem{Mohapatra:1982aq}
  R.~N.~Mohapatra and G.~Senjanovi\' c,
  ``Higgs Boson Effects in Grand Unified Theories,''
  Phys.\ Rev.\ D {\bf 27}, 1601 (1983).
 % doi:10.1103/PhysRevD.27.1601
  %%CITATION = doi:10.1103/PhysRevD.27.1601;%%
  %82 citations counted in INSPIRE as of 16 Nov 2016

%\cite{Jones:1981we}
\bibitem{Jones:1981we}
  D.~R.~T.~Jones,
  ``The Two Loop beta Function for a G(1) x G(2) Gauge Theory,''
  Phys.\ Rev.\ D {\bf 25}, 581 (1982).
  %%CITATION = PHRVA,D25,581;%%
  %231 citations counted in INSPIRE as of 25 août 2015

%\cite{Nishino:2012bnw}
\bibitem{Nishino:2012bnw}
  H.~Nishino {\it et al.} [Super-Kamiokande Collaboration],
  ``Search for Nucleon Decay into Charged Anti-lepton plus Meson in Super-Kamiokande I and II,''
  Phys.\ Rev.\ D {\bf 85}, 112001 (2012)
 % doi:10.1103/PhysRevD.85.112001
  [arXiv:1203.4030 [hep-ex]].
  %%CITATION = doi:10.1103/PhysRevD.85.112001;%%
  %100 citations counted in INSPIRE as of 16 Nov 2016




%\cite{Machacek:1979tx}
\bibitem{Machacek:1979tx}
  M.~Machacek,
  ``The Decay Modes of the Proton,''
  Nucl.\ Phys.\ B {\bf 159}, 37 (1979).
 % doi:10.1016/0550-3213(79)90325-0
  %%CITATION = doi:10.1016/0550-3213(79)90325-0;%%
  %127 citations counted in INSPIRE as of 12 Dec 2016



%\cite{Nath:2006ut}
\bibitem{Nath:2006ut}
  P.~Nath and P.~Fileviez Perez,
  ``Proton stability in grand unified theories, in strings and in branes,''
  Phys.\ Rept.\  {\bf 441}, 191 (2007)
  %doi:10.1016/j.physrep.2007.02.010
  [hep-ph/0601023].
  %%CITATION = doi:10.1016/j.physrep.2007.02.010;%%
  %251 citations counted in INSPIRE as of 09 Jun 2016


%\cite{Aoki:2013yxa}
\bibitem{Aoki:2013yxa}
  Y.~Aoki, E.~Shintani and A.~Soni,
  ``Proton decay matrix elements on the lattice,''
  Phys.\ Rev.\ D {\bf 89}, no. 1, 014505 (2014)
  %doi:10.1103/PhysRevD.89.014505
  [arXiv:1304.7424 [hep-lat]].
  %%CITATION = doi:10.1103/PhysRevD.89.014505;%%
  %37 citations counted in INSPIRE as of 06 Sep 2016





\end{thebibliography}
\end{document}